# Ultra-rapid, Quantitative, Label-free Antibiotic Susceptibility Testing via Optically Detected Purine Metabolites

A. Fraiman and L. D. Ziegler[†]*

Department of Chemistry and The Photonics Center, Boston University, Boston MA 02215

## Abstract

In order to facilitate the best antimicrobial prescribing practices and to help reduce the increasing global threat of antibiotic resistance, there is an urgent need for the development of novel and truly rapid (≤ 1 hour) antibiotic susceptibility testing (AST) platforms. A 785 nm surface enhanced Raman spectroscopy (SERS) based phenotypic methodology is described that results in accurate minimum inhibitory concentration (MIC) determinations for all tested strain/antibiotic pairs. The SERS-AST procedure results in accurate MICs in ~1 hour, including a 30-minute incubation period, and is effective for both Gram positive and negative species, and for antibiotics with different initial primary targets of antibiotic activity, and for both bactericidal and bacteriostatic antibiotics. The molecular level mechanism of this methodology is described. Bacterial SERS spectra are due to secreted purine nucleotide degradation products (principally adenine, guanine, xanthine and hypoxanthine) resulting from water washing induced bacterial stringent response and the resulting (p)ppGpp alarmone mediates nucleobase formation from unneeded *t*RNA and *r*RNA. The rewiring of metabolic responses resulting from the secondary metabolic effects of antibiotic exposure during the 30-minute incubation period accounts for the dose dependence of the SERS spectral intensities which are used to accurately yield the MIC. This is the fastest demonstrated AST method yielding MICs.

*Corresponding author: lziegler@bu.edu

[†]ORCID 0000-0002-3026-8986

# I. Introduction

Infectious diseases remain one of the greatest challenges to global health,[1] and the growing prevalence of infections resulting from drug-resistant bacteria, in particular, are increasingly recognized as a major public health problem of the 21st century.[2] In 2019, 1.27 million deaths world-wide were attributed to bacterial antimicrobial resistance (AMR)[2] and by 2050 it has been estimated that bacterial AMR could kill as many as 10 million people annually when it could potentially become the world's primary cause of death.[3] A multifaceted approach, including the development of new antibiotics, more judicious use of existing efficacious drugs, a better understanding of resistance acquisition mechanisms, and improved diagnostics is needed to address this worldwide problem.[4] The blanket use of empirical antibiotic before susceptibility testing results are determined, an almost standard current clinical practice, has further contributed to the exploding evolution of AMR. Thus, a key component of this needed concerted effort is for clinicians to have the known drug susceptibility profiles of the causative pathogen(s) in a fast enough time frame to reduce the unnecessary use of antibiotics altogether and promote the use of narrow spectrum drugs.[3]

In addition, delays in the start of the most effective antibiotic treatment has deleterious health outcomes resulting in higher morbidity and mortality rates, and accompanying consequent higher economic costs.[5, 6] Every *hour* delay in the administration of the correct antibiotic in the treatment of bacteremic patients, for example, increases the risk of mortality by 9%.[7] A novel and truly rapid AST platform that can yield results in a 30 minute to 1 hour timeframe has been explicitly identified as an urgent and aspirational need for the most effective management of infectious diseases,[8] and can also serve to lower the costs of much needed new drug development by facilitating clinical trials.[9]

Most current instrumentation providing quantitative drug susceptibility profiles in clinical settings, such as the biomerieux Vitek 2 or BD Phoenix, rely on automated bacterial growth-based measurements. While such instruments can provide antibiotic susceptibility testing (AST) results in 4 – 18 hours starting with clinical isolates, such traditional growth based AST approaches, even for rapidly growing bacterial species, typically take *a minimum* of two days to obtain susceptibility results from clinical samples including initial cultivation, isolation and enrichment steps.[10, 11]

Both genotypic and phenotypic methodologies have been proposed and recently demonstrated to reduce times for more targeted drug treatment decisions relative to current gold standard growth based methods.[8, 10, 12] Genotypic methods for identifying antimicrobial resistance rely on the direct or indirect detection of specific resistance genes[13] which can be used to predict phenotypic AST results and thus guide therapy selection. While genotypic methods are generally fast[14] (1 – 5 hour turnaround times) and effective at predicting some specific antimicrobial resistance, they do not provide quantitative antimicrobial susceptibility measures, such as the minimum inhibitory concentration (MIC). The MIC is the widely used quantitative measure of *in vitro* drug susceptibility corresponding to the smallest antibiotic concentration that will prevent the growth of bacterial cells in growth media. Furthermore, the ability to detect all resistant phenotypes is limited by the range of known resistance determinants, the lack of straightforward genotype identification and the ability to identify emerging resistant strains. Relative costs aside, the principal drawback to phenotypic AST is the slower time to result.

Phenotypic AST methodologies may be categorized into techniques that directly or indirectly detect measurements of growth metrics, such as doubling time or biomass change or, those that are based on non-cell growth bacterial responses allowing either quantitative (MIC) or categorical (S/R/I susceptible/resistant/intermediate) measures of a pathogen's susceptibility to antibiotic exposure. However, traditional and emerging AST methodologies that rely on cell growth responses face a fundamental limit for minimizing the time to results for AST.[15] Some examples of recent non-growth based techniques demonstrating rapid AST include measurement of AFM vibrations,[15] frequency dependent impedance cytometry,[16] and rapid metabolomic activity.[12, 17] The ultra-rapid, quantitative results described here belong to this last growth-free phenotypic AST category that exploits an optical approach for the detection of inherent, rapid bacterial stress responses to two environmental threats; antibiotics and starvation.

Raman scattering is a relatively fast, label-free optical technique that can identify molecular species due to their unique vibrational signatures. Additionally, like mass spectrometry, it is a multiplexing analytical technique capable of identifying multiple molecular components in, for example, complex biological samples, due to the narrow width of the spectral features. Attempts to exploit Raman spectroscopy for relatively rapid AST have been recently demonstrated.[12, 18, 19] In one class of efforts, changes in the relative intensity of bacterial spectral features following antibiotic exposure are taken as measures of antibiotic efficacy. However, these methods rely on detecting known resistance mechanisms and may not be applicable for the required wide range of bacterial species or antibiotics and unknown resistance, and how to achieve quantitative MIC determinations is not clear. In another approach, both spontaneous and stimulated Raman spectroscopy (SRS) have been used to determine drug susceptibilities from the appearance rates of C-D vibrational bands following $D_2O$ exposure in the absence and presence of antibiotics. MIC determinations via SRS require two incubation periods and sophisticated laser instrumentation for determining MICs in 2.5 hours via $D_2O$ uptake measurements.[20] The analogous $D_2O$ uptake MIC determinations via spontaneous Raman required 5 hours.[21]

Surface enhanced Raman spectroscopy (SERS) exploits plasmonic resonance effects arising from metal nanostructured surfaces to enhance Raman intensities.[22] Some molecules very close to nanostructured metal surfaces, typically Au or Ag, can exhibit Raman scattering intensities enhanced by $10^6 – 10^9$ when the incident and scattered light frequencies are coincident with the surface plasmon resonance of the nanostructured surface. In prior work we showed that the SERS spectra of viable bacterial cells excited with 785 nm radiation were exclusively due to secreted purines in the near cellular or extracellular region.[23-25] Characteristic, strain specific SERS spectra resulted from different amounts of these secreted purine components.[23-25] These purines, in turn, resulted from the nucleotide degradation process triggered by the bacterial stringent response (SR).[23] The SR (also referred to as the starvation response) is a broadly conserved bacterial stress response that controls adaptation to nutrient deprivation.[26-29] In this SERS context it is initiated when bacterial cells are washed in water to remove all growth media or residual body fluid components prior to SERS signal acquisition and is thus an inherent consequence of the required bacterial sample preparation protocol. Furthermore, SERS spectral acquisitions at the single bacterial cell level have been demonstrated, thus offering an approach with potentially high sensitivity.[30]

Here we show how SERS-based phenotypic AST profiles can be determined in one hour or less based on the relative SERS intensities of the secreted bacterial purine metabolic markers as a function of drug concentration exposure during incubation. New details on biochemical mechanisms connecting the SR and enhanced purine degradation, the detection step of this AST methodology, are discussed. Furthermore, recent progress in understanding how secondary metabolic bacterial responses play a crucial role in antibiotic responses are also summarized here in order to understand the observed dramatic dose-dependent effects on SR purine secretion levels in response to antibiotics.

The use of SERS intensities for determining drug susceptibility profiles has been demonstrated by one other research group.[31, 32] However, here we emphasize the molecular level analysis of these spectra and their biochemical origins, essential for optimizing this SERS-based approach, as well as demonstrating important differences, including time-to-results, that distinguishes our approach and its improved performance characteristics.

## II. Materials and Methods

**1. Bacterial samples and SERS AST procedure.** Bacterial strains used in this study were purchased from ATCC or were clinical isolates donated by BD Life Sciences.[24] The bench top procedure resulting in the SERS determinations of MIC values for the strain/drug pairs discussed here is outlined in Fig. 1. A range of doubling concentrations of the selected antibiotic (~ 0, x, 2x, 4x, 8x, 16x, etc.) are added to 10 mL of Mueller Hinton Broth (MHB) (Sigma) in 50 mL tubes and warmed to 37 $^{o}$C. Overnight bacterial culture was inoculated into ~10 mL of MHB and the bacterial concentration was adjusted to yield an OD ~ 0.1 at 600 nm or an initial concentration on the order of ~$10^6$ cfu/mL. 0.25 mL of the ~0.1 OD bacterial solution is added to each tube which thus contains the same number of bacteria (~2.5 x $10^5$) with different antibiotic concentrations. The range of antibiotic concentrations is intended to encompass the MIC in addition to including a self-calibrating sample without any antibiotics. These culture samples are immediately incubated for 30 min at 37 $^{o}$C with shaking. After this bacterial incubation period with growth media and specified antibiotic concentration, the cultures are transferred into 2 mL μ-centrifuge tubes via centrifugation to remove MHB. For the data reported here 4-5 aliquots from the 10 mL culture tube were sequentially added to the smaller tubes and following each successive centrifugation (1 minute, 12,000 RPM, 20C), the excess both was removed and the cell pellet size increased. The remaining cell pellets were washed with 1 mL of pH 7 water three times. Before centrifugation the aqueous solution was vortexed (~10 sec) to ensure good water mixing and effective water washing. Following removal of the supernatant after the last wash, ~5 μL remains in the centrifugation tube. The remaining sample is vortexed one more time and a 1 μL pipet is used to transfer a portion of the resulting suspension onto the SERS chip and warmed at ~37 $^{o}$C to dry. SERS measurements were made after ~5 minutes when nearly all the water on the SERS substrate had evaporated. The entire process including the 30 minute incubation step takes ~1 hour for a given bacteria/antibiotic combination.

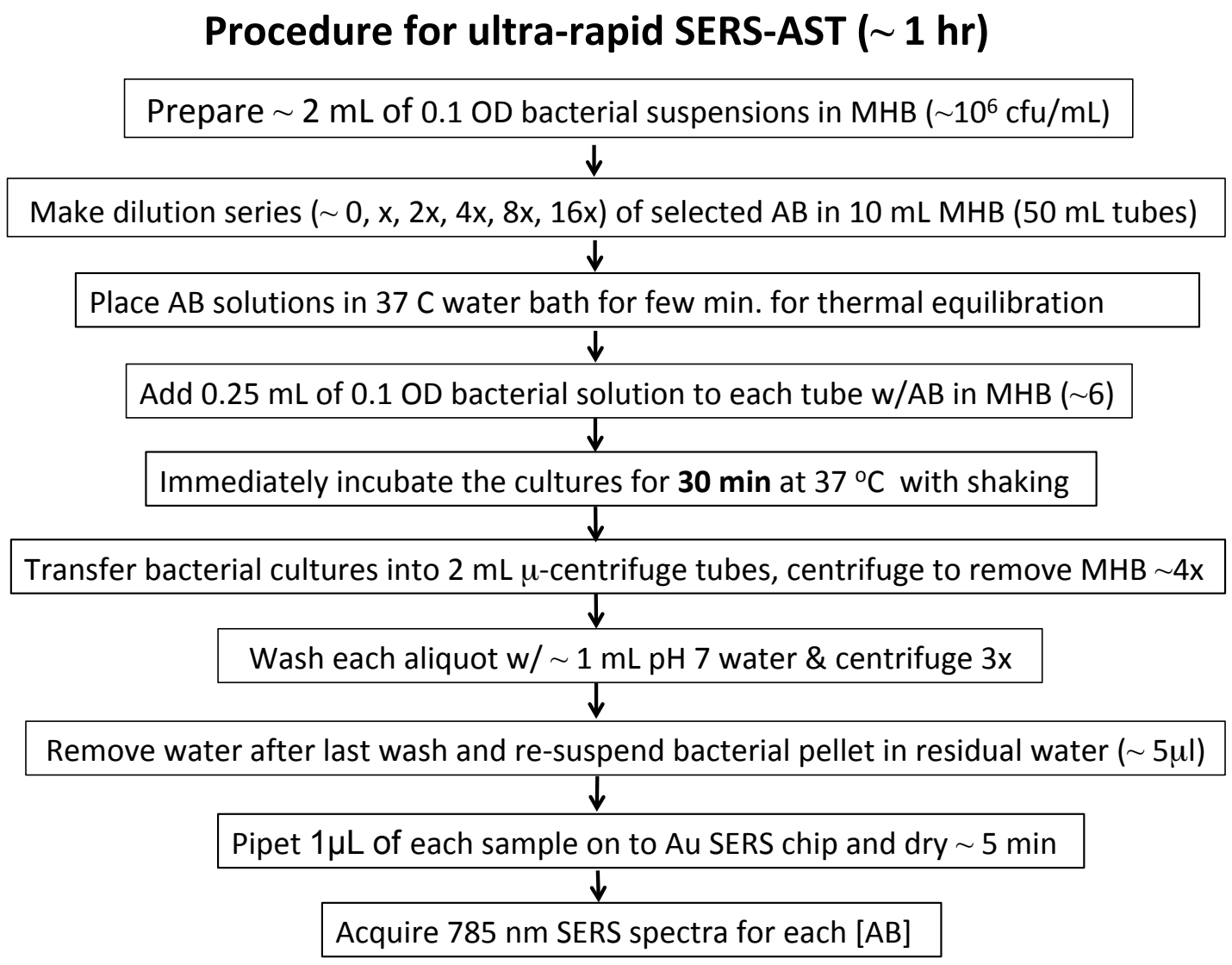

**Figure 1.** . Outline of the procedure used to acquire the ultra-rapid SERS-AST results on Au nanoparticle substrates.

**2. SERS substrates.** All SERS spectra reported here were obtained using *in-situ* grown, aggregated Au nanoparticle covered $SiO_2$ substrates developed in our laboratory.[30] Details concerning the production of these SERS active chips and the characterization of their performance for providing reproducible SERS spectra of bacteria have been described previously. In brief, these substrates are produced by a two stage reduction of an Au ion doped sol-gel and results in small (2 – 15 particles) aggregates of monodispersed ~80 nm Au nanoparticles covering the outer layer of ~1mm$^2$ $SiO_2$ substrate.[30]

**3. SERS Spectral acquisition**. 785 nm excited bacterial SERS spectra were acquired with an RM-2000 Renishaw Raman microscope employing a 50x objective. Incident laser power of ~1 mW and ~10 seconds of illumination time were used to collect single spectra. The observed spectra typically resulted from ~10 bacterial cells within the field of view (~100 μm$^2$) and are from the same biological samples that were used to determine the broth dilution MIC results. The displayed spectra are the average of 5 – 6 spectral acquisitions for each antibiotic concentration at different locations on the SERS chip. As demonstrated previously, spectral standard deviations for bacterial SERS spectra with excitation conditions and sample concentrations comparable to those used here are shown to be ±15% across the entire spectrum.[24, 25] The 520 cm$^{-1}$ band of a silicon wafer was used for frequency calibration. Peak frequency precision is ± 0.5 cm$^{-1}$.

**4. Data fitting analysis.** GRAMS/AI™ Spectroscopy Software was used to manually baseline correct the experimental SERS spectra. Averaged bacterial spectra were empirically best-fit by adjusting component contributions from normalized purine SERS spectra. Normalized SERS spectra of ~20 μM aqueous solutions of the purine components were independently obtained and used for this best-fitting purpose. Due to the broad SERS spectral baseline variability and some systematic vibrational frequency shifts between the bacterial and purine only solutions, excellent best fits to the observed bacterial spectra could be achieved by an empirical fitting procedure as demonstrated previously.[23]

**5. Evaluation of Antibacterial Activity via Growth.** Reference MIC values were determined using a broth dilution method according to CLSI guidelines.[33] Bacterial samples in MHB mixed with doubling concentrations of antibiotics were incubated at 37 °C for 24 hours and growth was assessed by visual inspection of turbidity or by optical extinction (OD) measurements at 600 nm. The reported gold standard MIC values corresponded to the lowest antibiotic concentration where no bacterial growth was observed; OD ~0. (See Fig. S1 for examples of the visual inspection results.) Antibiotics used in this study were chosen in part to represent the major classes of initial bacterial interaction mechanisms (see Table 1) and to contrast the SERS response of different strains to the same drug or of the same strain to different drugs.

## III. Results

### A. Quantitative, ultra-rapid MIC via SERS

A detailed phenomenological example of how SERS provides ultra-rapid (~1 hour) drug susceptibility information is illustrated in Fig. 2 for *E. coli* ATCC 2452. Following the gold standard CLSI growth-based procedure for determining quantitative drug susceptibilities,[33] this Gram-negative strain is found to be susceptible to the broad spectrum bacteriostatic antibiotic, tetracycline, but resistant to the $\beta$-lactam, ampicillin. Optical density (OD) turbidity measurements at 600 nm after a 24 hour growth period as a function of doubling tetracycline and ampicillin concentrations in the incubated growth media pathogen solution are given in Figs. 2a and 2d. The lowest drug concentration where the bacterial/drug/media solution shows an OD of ~0 after 24 hours is the MIC value via this traditional growth procedure. Thus, *E. coli* 2452 exhibits a MIC of 2 mg/L due to tetracycline (Fig. 2a) via 24 hour growth measurements. In contrast, incubation with ampicillin doses up to 128 mg/L (Fig. 2d) has no discernable effect on the growth of this strain after 24 hours as evidenced by the unchanged OD for all *E. coli* 2452 samples with ampicillin concentrations in this range and is thus ampicillin resistant.

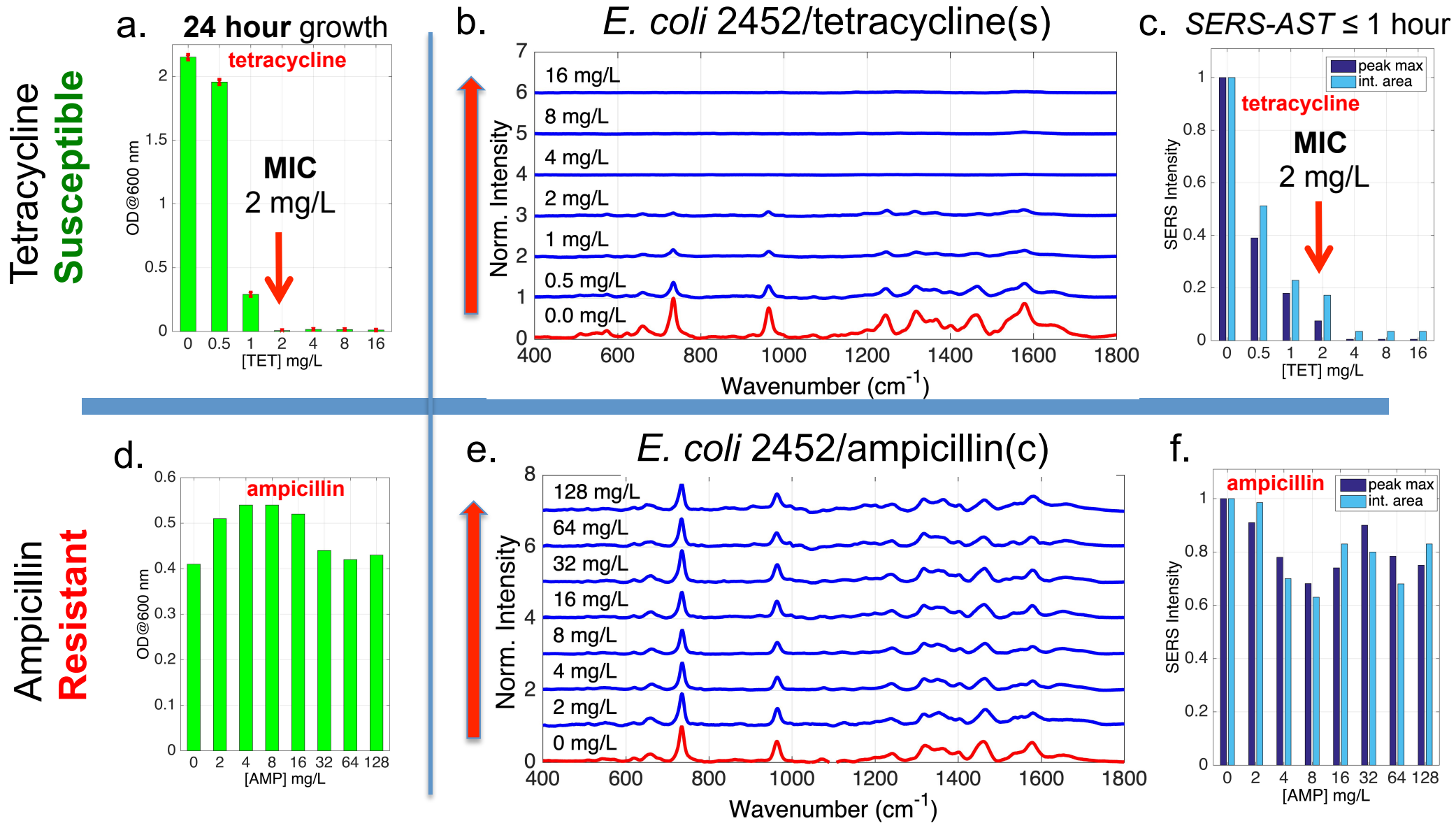

**Figure 2.** SERS-AST results for a single *E. coli* strain and two different antibiotics a. OD measures determine that the MIC for *E. coli* 2452 is 2 mg/L of tetracycline after 24 hours via broth dilution. b. SERS spectra of *E. coli* 2452 after 30-minute incubation with tetracycline and growth media as a function of doubling antibiotic concentration. c. *E. coli* 2452 SERS intensity measured by the spectral peak maximum (~730 $cm^{-1}$) and integrated spectral area as a function of tetracycline concentration. SERS-AST determined MIC given by antibiotic concentration when the intensity first falls below 20% of the maximum (see discussion in text). d. – f. are the corresponding growth determined MIC, SERS spectra and representations of the SERS intensity for the *E. coli* 2452 strain incubated with ampicillin. The SERS intensity independence indicates this strain is resistant to ampicillin as found both by the 24-hour growth (d.) and ~one hour SERS AST approaches (e., f.). The (c) and (s) designations indicate if the antibiotic is bactericidal or bacteriostatic.

The corresponding 785 nm excited SERS spectra of *E. coli* 2452 as a function of doubling tetracycline concentrations during a 30-minute incubation period are displayed in Fig. 2b. Following this incubation in MHB growth media and a specific antibiotic concentration, bacterial solutions are water washed and centrifuged three times before a 1 μL bacterial solution is dropped on the Au SERS substrate a Raman signal is acquired, as described above. A SERS spectral measurement is completed within one hour including the 30-minute drug incubation, washing and SERS signal acquisition. The spectra shown in Fig. 2b are all normalized to the maximum intensity of the most intense SERS spectral feature observed for this series of *E. coli* 2452/tetracycline spectra which is the ~730 $cm^{-1}$ band in the *E. coli* sample lacking any tetracycline (0 g/mL) during the 30 minute incubation period. The relative intensities of these SERS spectra decrease as a function of tetracycline concentration over the 0 to 16 mg/L range (Fig. 2b). To quantify the effect of antibiotic exposure on these *E. coli* 2452 SERS spectra, both the peak intensity of the ~730 $cm^{-1}$ band (dark blue bars) and the total (400 – 1800 $cm^{-1}$) integrated SERS spectral intensities (light blue bars) are plotted as a function of tetracycline concentration during incubation in Fig. 2c. The same trend of monotonically decreasing SERS intensity as a function of antibiotic exposure is observed whether the maximum peak or total integrated intensity is used as the figure of merit to quantify the effect of tetracycline on the relative *E. coli* 2452 SERS intensities.

In contrast, the *E. coli* 2452/ampicillin spectra (Fig. 2e) measured after the 30-minute incubation in MHB containing doubling concentrations of ampicillin exhibit no substantive change in intensity as a function of ampicillin concentration (up to 128 mg/L) within experimental uncertainty (± 15%). More quantitatively, bar plots of the relative maximum peak intensity at ~730 $cm^{-1}$ (dark blue bar) or the integrated area (light blue bar) of these *E. coli* 2452 SERS spectra (Fig. 2f) illustrate this ampicillin concentration independence. Incubation with 16 mg/L of tetracycline results in a more than 200 fold reduction in *E. coli* 2452 SERS peak intensity relative to no drug exposure, whereas there is virtually no change in the SERS intensity of this strain at the same ampicillin concentration. The effectively unchanged SERS intensities following incubation with ampicillin are consistent with the resistance of this strain as determined by broth dilution results (Fig. 2d).

Following the sample preparation and experimental procedures employed here, we find that to achieve, essential agreement with 24 hour growth based results, the MIC determined via this AST-SERS methodology is given by the antibiotic concentration where the normalized SERS maximum peak or integrated area intensity falls below 18% of the maximum SERS intensity. For the *E. coli* 2452/tetracycline combination the SERS-based determined MIC is thus 2 mg/L (Figs. 2a and 2c). While this empirically-determined cutoff consistently results in MIC values that match the gold standard growth MIC within ±1 doubling concentration (essential agreement[34]) for all susceptible bacteria-drug combinations we have tested (*vide infra*), the SERS-based MIC is determined about twenty times faster (~1 hour) than the CLSI gold standard growth based result for a given bacterium/drug concentration. Although there is no *a priori* fundamental significance to this ~18% of maximum SERS cutoff value yielding MIC determinations, we have phenomenologically found that a > 80% drop in SERS intensity (peak or integrated) consistently results in essential agreement between the SERS-AST results and the 24 hour gold standard MIC determinations (see further results below). Although the growth turbidity OD value goes to ~0 at the MIC, there is no reason for the SERS-AST MIC to be given by a null result because the SERS MIC measure is the result of a completely different biochemical mechanism than bacterial cell growth suppression, as discussed below.

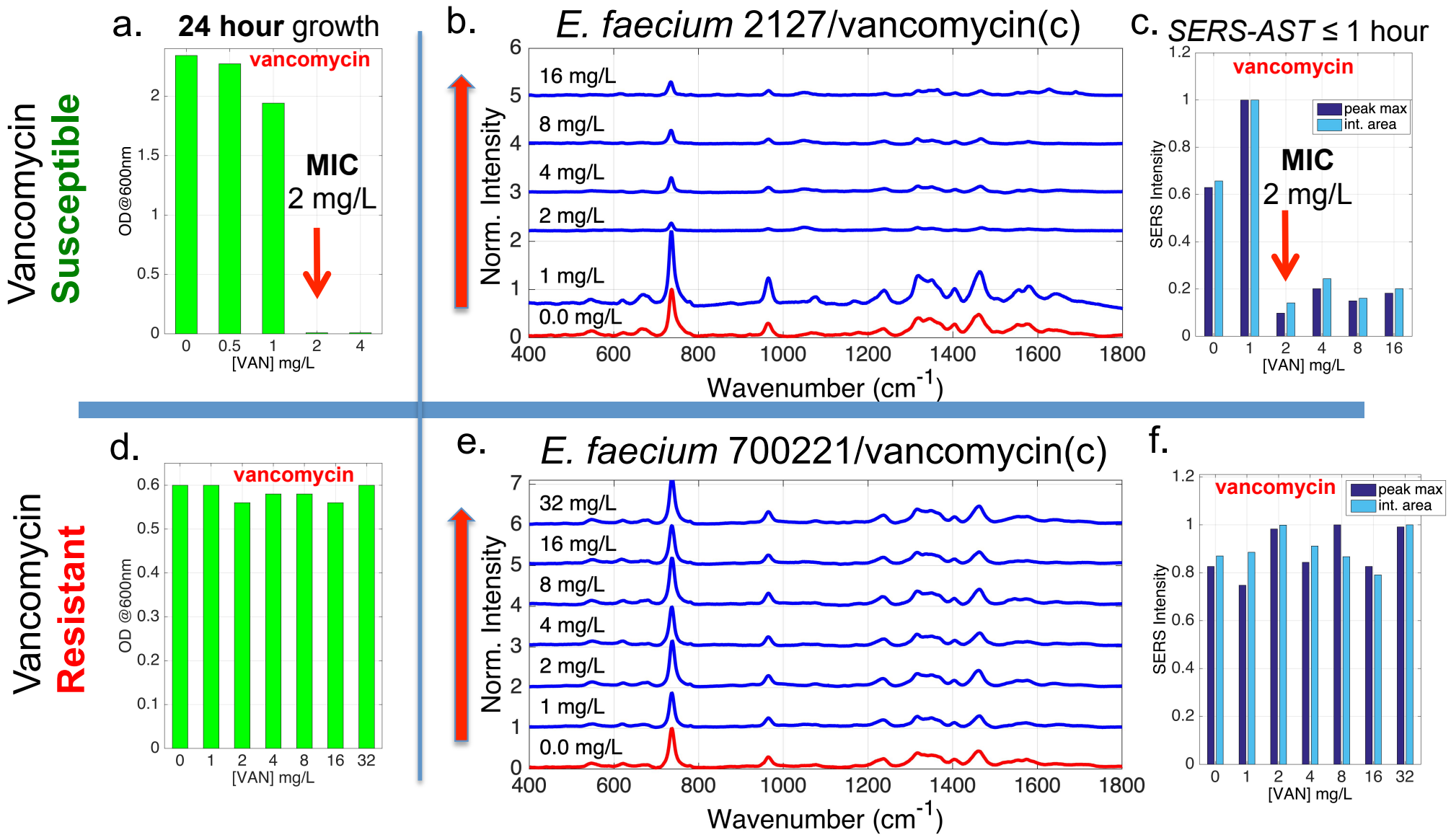

**Figure 3.** SERS-AST results for two *E. faecium* strains and vancomycin. a. The MIC is 2 mg/L for *E. faecium* 2127 dosed with vancomycin via broth dilution methodology. b. SERS spectra of *E. faecium* 2127 after a 30-minute incubation with vancomycin and MHB as a function of doubling antibiotic concentration. c. *E. faecium* 2127 SERS intensity as given by the normalized spectral peak maximum (~730 $cm^{-1}$) and integrated entire spectral area as a function of [vancomycin]. SERS-AST determinations take ~one hour including incubation time. The SERS-AST MIC value given by drug concentration where the intensity first falls below 20% of the maximum (see discussion in text). d. – f. are the corresponding growth determined MIC, SERS spectra and corresponding SERS intensity bar plots for *E. faecium* 700221 incubated with vancomycin. The concentration independence indicates this strain is resistant to vancomycin as found by both the 24-hour growth (d.) and ~1- hour SERS AST approach (e., f.). The (c) designation indicates vancomycin is a bactericidal antibiotic.

In contrast to Fig. 2, the rapid SERS and 24-hour growth responses of two strains of the same species with different susceptibility/resistance responses to the same antibiotic is demonstrated in Fig. 3. *E. faecium 2127* and *E. faecium 700221* are Gram-positive enterococcus strains that respectively lack and possess vancomycin resistance genes. By broth dilution *E. faecium* 2127 exhibits a vancomycin MIC of 2 mg/L (Fig. 3a) and *E. faecium* 700221 exhibits vancomycin resistance, at least up to 32 mg/L (Fig. 3d) in accordance with their genetic markers. The corresponding SERS spectra of these two *E. faecium* strains as a function of vancomycin concentration exposure during the 30-minute incubation are displayed in Figs. 3b and 3e. The SERS intensity of the vancomycin resistant strain, *E. faecium* 700221, is effectively independent of the drug concentration (Fig. 3e) by both spectral peak maximum (~730 $cm^{-1}$) and integrated area measures within experimental uncertainty (Fig. 3f). The SERS peak and integrated intensity of *E. faecium* 2127 spectra, in contrast, first increases at the lowest vancomycin exposure concentration (1 mg/L) and then dramatically decreases with higher concentrations (Fig. 3c). The SERS responses of these two *E. faecium* strains are consistent with their 24 hour growth vancomycin susceptibility and resistance character. The relative SERS intensities of the *E. faecium* 2127 spectra measured by either relative spectral peak maximum intensity or integrated area falls to < 18% of its peak value for 2 mg/L (Fig. 3b, c) thus establishing this value as the SERS-AST MIC in agreement with the 24 hour growth results (Fig. 3a). Again, the drastically reduced time to result (~1 hour for the SERS determination) is the only difference between the outcomes of the growth and SERS AST results.

Bar graphs summarizing another thirteen examples of this novel, rapid, phenotypic SERS-AST methodology for quantitative MIC determinations are shown in Figs. 4 and 5 to further establish the general applicability and accuracy of this approach. The same experimental

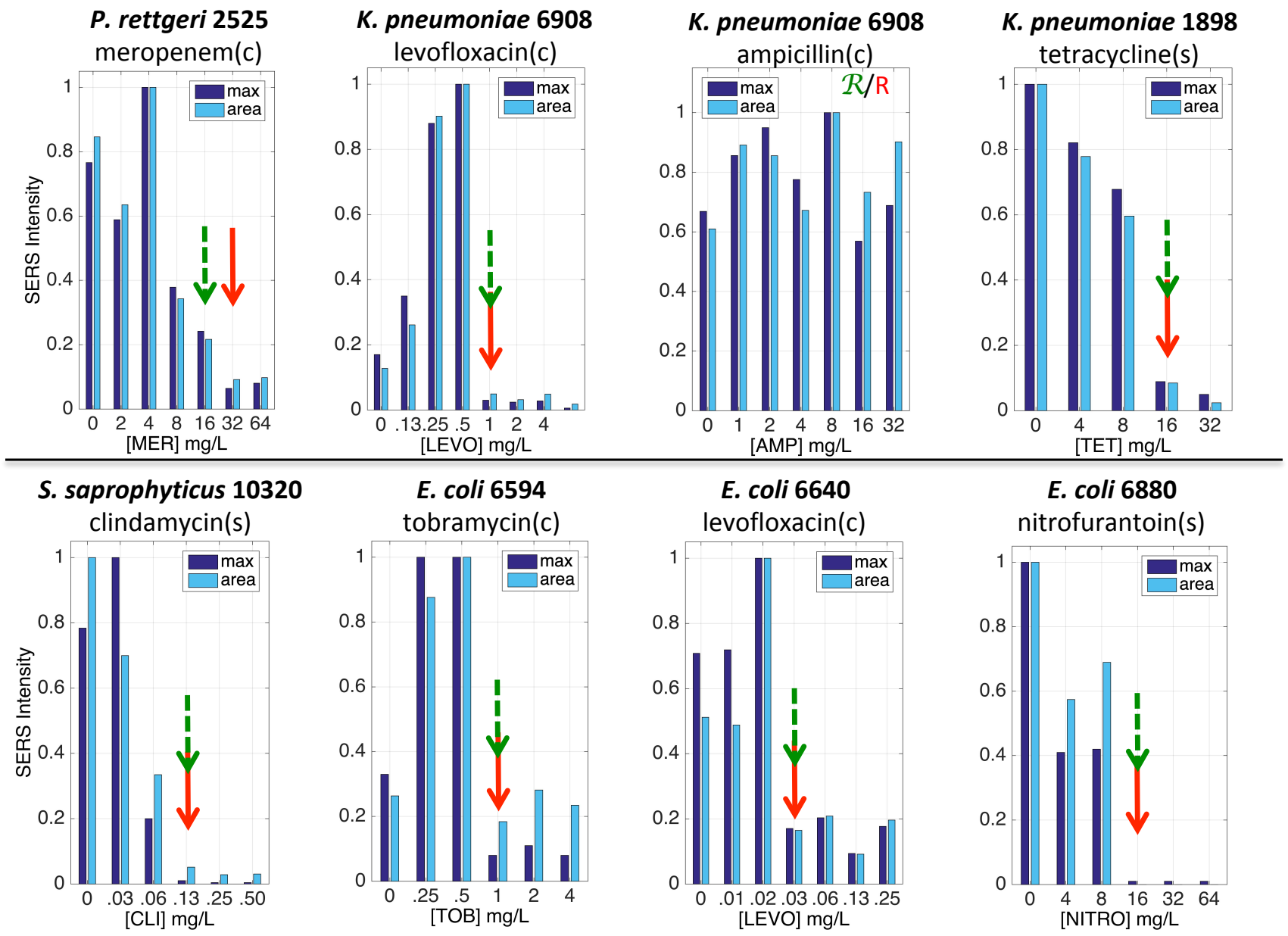

Figure 4. Additional examples of rapid SERS-AST determined MICs. Bar graphs represent the normalized integrated area (light blue) or peak maximum (dark blue) of the SERS spectra of eight bacterial species after 30-minute incubation with the indicated antibiotic concentration as shown in Figs. 2c, 2f and 3c, 3f. The dashed green and solid red arrows are the 24 hour growth, determined by visual inspection, and corresponding SERS determined MICs, respectively. *K. pneumonia* 6908 is found to be ampicillin resistant by both methods. The (c) and (s) designations correspondingly indicate if the antibiotic is bactericidal or bacteriostatic.

protocols (Sec. II) used for acquisition of the data in Figs. 2 and 3, i.e. 30 min incubation with antibiotics and subsequent water washing for a total AST time ≤ 1 hour, were followed. The dose dependent SERS spectral intensities represented by these bar graphs are given in Supplementary Information (Figs. S2-S14). Each of these SERS bacterial strain/drug AST profiles, i.e. SERS intensity per doubling antibiotic concentration (Figs. 4 and 5), is given for both the maximum SERS peak intensity and the total integrated area. The dashed vertical green arrow in each Figs. 4 and 5 panel indicates the 24 hour growth MIC determined by visual

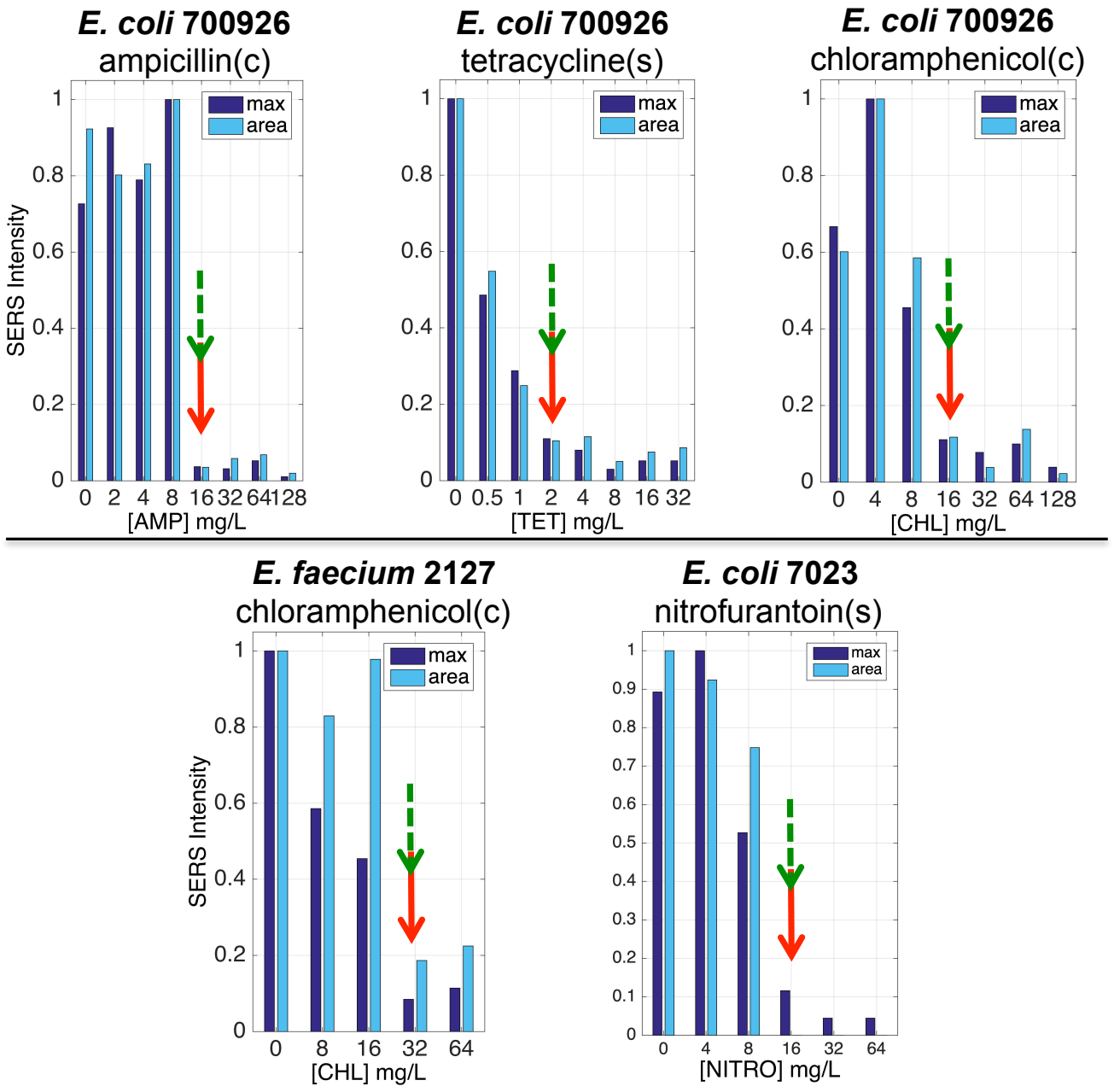

**Figure 5.** Additional demonstrations of the rapid SERS-AST methodology. Bar graphs represent the integrated area (light blue) or peak maximum (dark blue) of the SERS spectra of five bacterial species after 30 minute incubation with the indicated antibiotic doubling concentration. The dashed green arrow is the 24 hour growth, determined by visual inspection, MIC and the solid red arrow is the corresponding SERS determined MIC. The (c) and (s) designations indicate if the antibiotic is bactericidal or bacteriostatic respectively

inspection, i.e. the lowest antibiotic concentration where the bacterial solution with MHB and antibiotics remain transparent after 24 hours (e.g. see Fig. S1). The SERS MIC value is indicated by the solid vertical red arrow in each bar graph panel and corresponds to the lowest concentration where the SERS intensity is ≤ 18% of the relative peak height or integrated spectral area maximum. While SERS intensity decreases could be observed as early as 10 minutes after antibiotic exposure (Fig. S15) for some strains, consistently accurate AST results for all tested strains is achieved after the 30-minute incubation period.

These results further demonstrate that the SERS determined MICs agree with the 24-hour growth values by the procedure described here. In fact, SERS-AST MICs are exactly identical with gold standard results except for the *P. reggetti* 2525/merepenum pair where the SERS MIC is found to be 1 doubling concentration different than the growth value which is within the accepted essential agreement range.[34] The ampicillin resistant *K. pneumoniae* 6908 strain shows no significant SERS intensity decrease, within experimental uncertainty, as a function of ampicillin concentration as observed for the previously described resistant *E. coli* and *E. faecium* strains (Figs. 2 and 3). However, when *K. pneumonia* 6908 is incubated with levofloxacin, a precipitous drop in SERS intensity is found at the growth determined MIC dose (Fig. 4). The fact that MICs determined by SERS spectral maxima (typically ~730 $cm^{-1}$) or the total integrated area are the same may have implications for instrumentation design exploiting this discovery for a rapid AST device.

Phenomenologically, SERS-AST profiles exhibit two general types of antibiotic concentration trends. In one, SERS intensities monotonically decrease with increasing drug concentration (e.g. *E. coli* 700926/tetracycline). For the other, SERS intensity increases greater than the 0 mg/L spectrum at sub-inhibitory low dose concentrations (e.g. *E. coli* 700926/ampicillin) before plummeting at higher doses. These observed concentration trends will be described further below.

While the SERS-AST profile trends of the peak and integrated area as a function of antibiotic concentration are the same within experimental precision, two exceptions within this group are observed. The susceptibility profile for the *S. saprophyticus* 10320/clindamycin pair (Fig. 4) via integrated area monotonically decreases as a function of clindamycin concentration, whereas the peak maximum first increases at the lowest concentrations before monotonically decreasing at higher clindamycin concentrations. Additionally, the *E. faecium* 2127 SERS peak intensity monotonically decreases with chloramphenicol concentration (Fig. 4) but the integrated SERS spectrum intensity remains high at lower concentrations before decreasing near the MIC. However, most importantly, the same MIC value is found by both SERS intensity measures in these data sets. The molecular origins accounting for these seemingly anomalous, non-monotonic SERS intensity trends are also addressed below.

**B. Observed generality of SERS-AST capability**

In this initial effort to establish the applicability of this SERS-AST approach for potential development as a general diagnostic platform we highlight that the SERS susceptibility profiles shown in Figs. 2 - 5 provide ultra-rapid and accurate MICs for both Gram-positive (e.g. *S. saprophyticus*, *E. faecium*) and Gram-negative (e.g. *E. coli*, *K. pneumonia, P. rettgeri*) bacteria following a 30-minute incubation period. Thus, the SERS-AST methodology is independent of bacterial cell wall structure. Secondly, there are a handful of main categories of antibiotic initial activity targets including molecules that inhibit cell wall or protein synthesis (via ribosomal disruption), or that interfere with DNA replication or folic acid metabolism.[35] Table I lists the antibiotics used in these AST studies and their corresponding initial mechanism of action which covers these major classes of antibiotic activity. The demonstrated results show that the SERS-AST methodology works for all these initial target classes of antibiotic-bacterial cell interactions.

In addition, drugs exhibiting antibiotic effectiveness may be classified into two groups depending on their bacterial cell effects. Bactericidal antibiotics result in bacterial cell death, while bacteriostatics prevent the growth of bacterial cells.[36] Table I also indicates this activity classification for the antibiotics tested in these studies and are correspondingly labeled with their cidal (*c*) or static (*s*) description in each displayed AST profile (Figs. 2 – 5). As seen, SERS-AST accurately and rapidly determines MIC values for both classes of antibiotic activity. Interestingly, there appears to be a strong correlation between the above noted SERS intensity increase/decrease at sub-inhibitory doses, especially for the total SERS intensity as given by integrated area, and the bactericidal/bacteriostatic antibiotic classification (Figs. 2 - 5) and is discussed further below.

## C. Molecular origins of SERS bacterial signals

A further demonstration[23-25] of the purine molecular components responsible for these 785 nm SERS spectra on Au substrates is shown in Fig. 6 for four representative strains in the absence of antibiotics. The nearly overlapping red and black spectra in each of the Fig. 6 panels are the molecular component best-fits (by visual inspection) and observed SERS spectra respectively. The "best-fit" spectra result from linear combinations of normalized 785 nm SERS spectra of the indicated purine molecular components alone. Even given potential errors due to spectral baseline determinations, the implicit assumption that the observed SERS signals are not saturated, and neglecting any inter-purine base pairing interactions effects, nearly all the observed vibrational frequencies and their relative intensities are captured by this fitting procedure. These results, also demonstrated in prior studies from this lab,[23-25] provide additional evidence that the major molecular components contributing to these SERS spectra are nearly exclusively purines and hence these are the molecular species responsible for providing the rapid bacterial SERS-AST response to drug exposure.

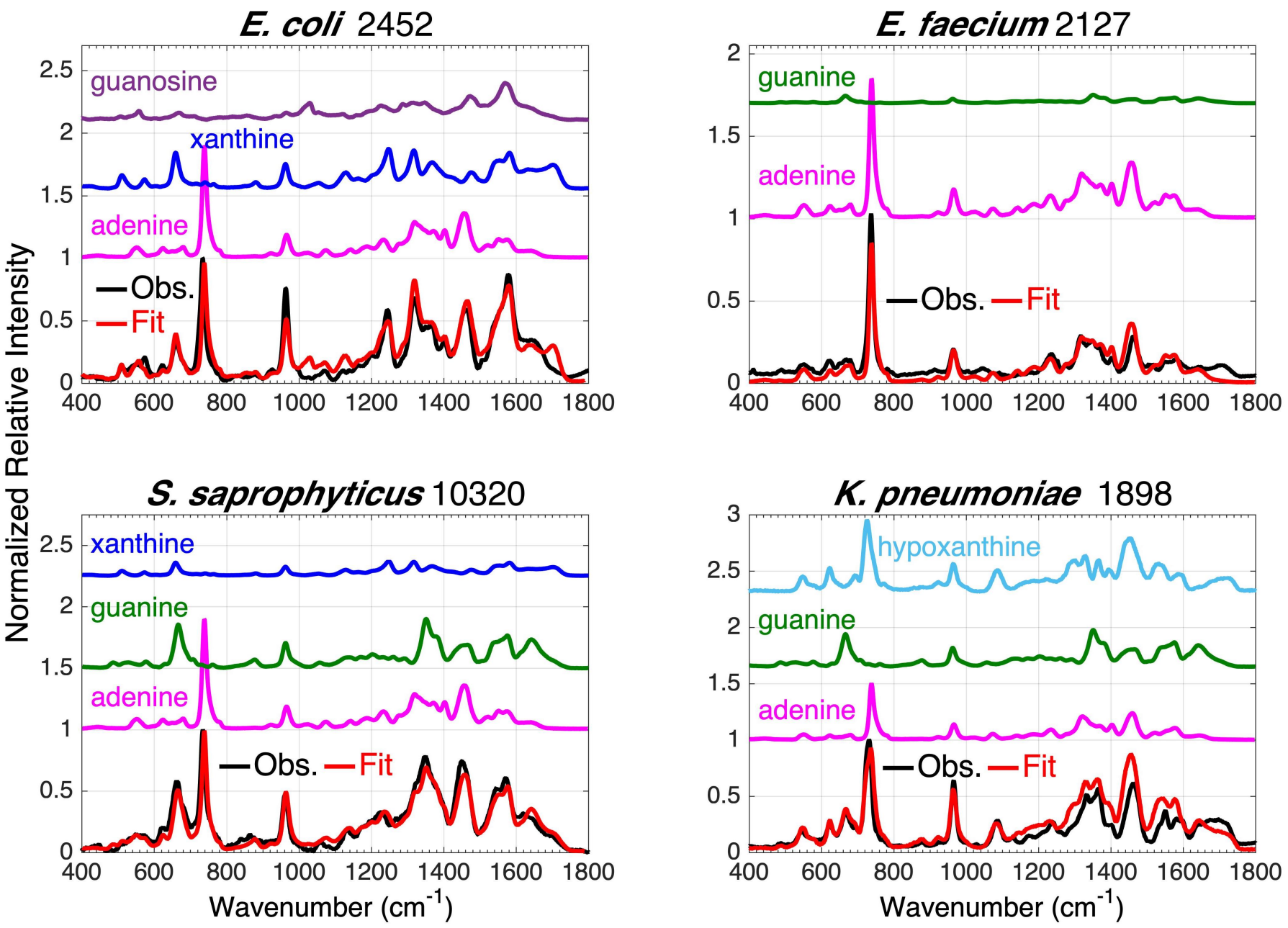

Figure 6. Empirically determined best-fits (red) of four representative normalized observed bacterial spectra (black) to a linear combinations of purine (adenine, hypoxanthine, xanthine, guanine, and guanosine) SERS spectra. Purines identified in SERS spectra of other strains include uric acid and adenosine. The relative contribution of each purine component to the total normalized bacterial SERS spectrum is shown in each displayed best-fit. This procedure identifies the molecular origins of nearly every vibrational band peak in these bacterial SERS spectra.

As shown in Fig. 6, the dominant molecular components to these SERS bacterial spectra are the purinergic nucleic acid degradation metabolites: adenine, hypoxanthine, guanine, xanthine, uric acid, guanosine, and adenosine. The SERS spectra of these purines given in each of the panels are scaled by their relative contribution to the total modeled bacterial spectrum. The different relative intensities, i.e. relative concentrations, of these compounds accounts for the spectral differences between bacterial species/strains, and potentially as a function of antibiotic exposure. For example, the SERS spectrum of *E. faecium* 2127 is dominated by adenine (90%) with just a small contribution (~5%) from xanthine. In contrast, the *E. coli* 2452 SERS spectrum is predominantly due to adenine (~60%), xanthine (~20%) and guanosine (~20%). Hence these SERS signatures have the potential capacity for bacterial identification within relatively limited reference SERS libraries, as previously demonstrated.[24, 37] Although nearly completely dominated by purine nucleic acid metabolites, relatively small amounts of the pyrimidine nucleobase cytosine were detected in a few bacterial spectra (e.g. *P. rettgeri* 2525, *E. coli* 700926*)*. The above indicated % compositions correspond to the relative amplitude of normalized purine SERS spectra and have not been corrected for relative SERS cross-section which would be indicative of the relative number density of these species. However, our principal goal here is to underscore the identity of the molecular species giving rise to these 785 nm excited SERS bacterial spectra[23-25] since the molecular origins of these SERS signals is central for understanding the biochemical basis for the reported ultra-rapid SERS-AST methodology.

**D. Bacterial SERS composition dependence on antibiotic dose**

Although the total integrated or peak SERS spectral intensities as a function of 30-minute antibiotic dose exposure provide the same MIC value, purine specific decomposition of the observed spectra as a function of drug dose during incubation allows a more quantitative assessment of the effects of antibiotic exposure pre-conditioning on the relative intensities for a given bacterial/drug pair. Such a concentration dependence may be anticipated if the purine contributions providing the AST information have metabolic pathway origins[23] and could play a role in analyzing some details of SERS-AST drug dependence. This molecular specificity also highlights how SERS may be used for studying the increasingly recognized role of purine biosynthesis in antibiotic efficacy.[38]

The molecular components of SERS spectra of six representative bacterial strain/drug pairs as a function of doubling antibiotic concentration are given in Fig. 7. The relative purine intensity contributions result from the best-fit procedure illustrated in Fig. 6. The relative purine composition (estimated uncertainty ±15%) is given for spectra *normalized* to maximum intensity at each drug concentration to highlight how the *relative* components change as a function of antibiotic dose. The MIC concentration is indicated in red along the drug concentration axis.

No single consistent purine composition trend with antibiotic concentration exposure is observed for these dose dependent SERS spectra. The relative contributions of the component purines to the SERS spectrum of susceptible strains generally exhibit incremental changes with incubated drug concentration (Fig. 7a-d). For example, for *K. pneumonia* 1898 (Fig. 7a), the relative concentration of hypoxanthine increases until the MIC in tetracycline (16 mg/L) is reached, and then decreases with increasing dose (Fig. 7a). The opposite trend is observed for the relative adenine contribution to the SERS spectra as a function of tetracycline for this bacteriostatic antibiotic. Furthermore, the relative guanine and xanthine contributions also exhibit opposite decreasing and increasing trends respectively for *K. pneumonia* 1898 responding to increasing tetracycline dose exposure. Similarly, gradual, changes in the relative purine contributions to the SERS spectra of *E. coli* 2452, *E. faecium* 2127 and *E. coli* 6640 to tetracycline, chloramphenicol and levofloxoacin respectively are found as shown in Figs. 7b, 7c and 7d. (Accurate relative purine analyses for *E. coli* 2452 at tetracycline concentrations greater than the MIC were not possible due to the weakness of these > MIC spectra. See Fig. 2b.) Interestingly, large changes in the chemical composition of the SERS spectrum do not consistently appear at the MIC as might have been anticipated, and the purine relative composition changes are generally systematic and incremental. On the other hand, the SERS spectral signature at the highest drug concentrations can be different than their corresponding drug-free SERS spectra as evident for the four susceptible strain/antibiotic combinations (Figs 7a-d). However, the AST profiles (Figs. 2 - 5) show that MIC determinations based on SERS spectral peak maxima or integrated spectral intensities are not affected by these dose dependent changes to the SERS spectral composition. Some disagreement between SERS-AST concentration profiles given by spectral maxima and integrated area can be attributed to changes in relative purine composition as a function of antibiotic concentration as seen for *S. saprophyticus* 10320/clindamycin and *E. facium* 2127/chloramphenicol (Fig. 5), although this does not hinder the accuracy of the SERS-AST MIC determination as these examples show because the overall intensity changes are the much larger dose dependent effect.

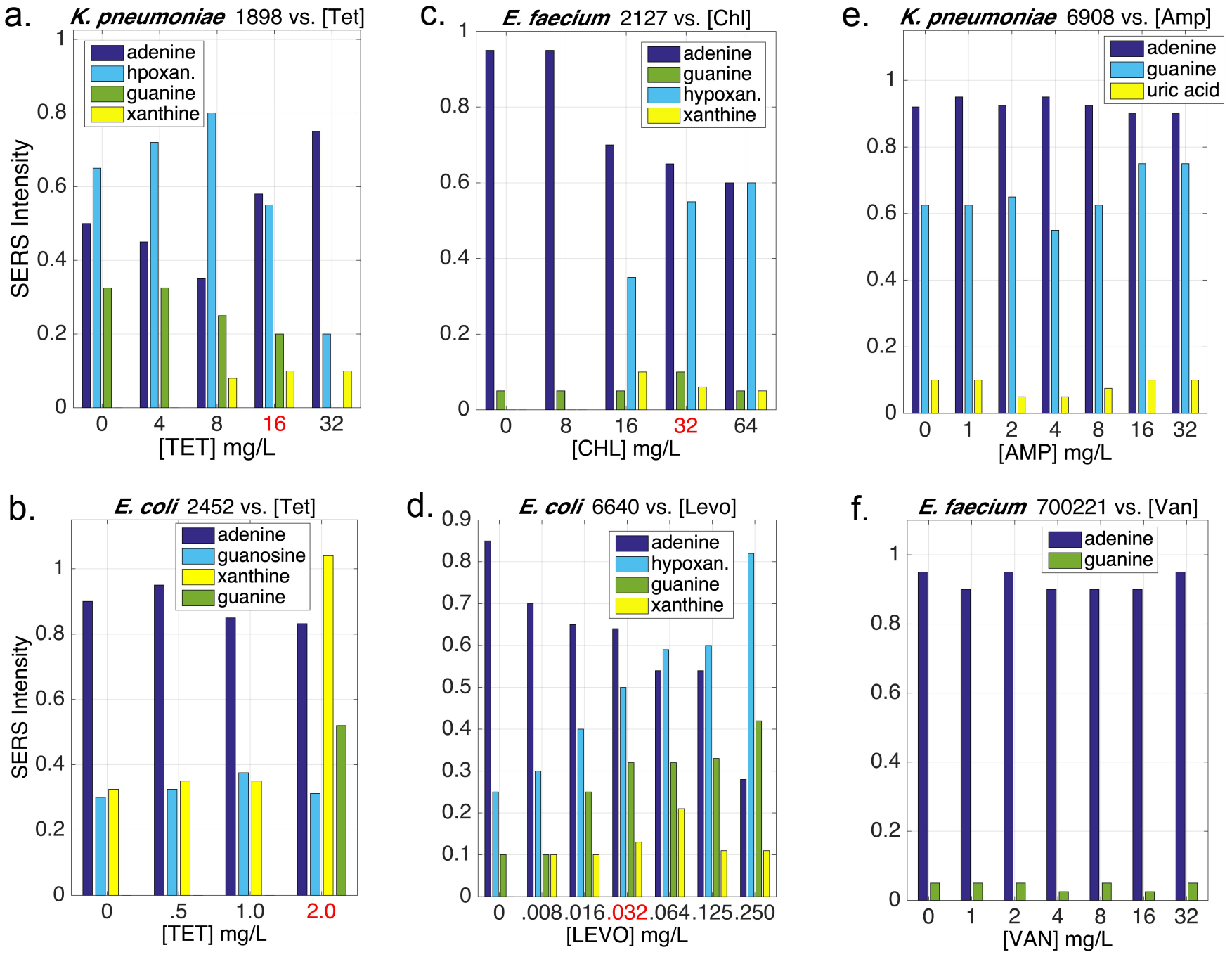

**Figure 7**. Dose dependence of the purine components of the normalized 785 nm SERS spectra for six representative bacterial strain/antibiotic combinations. The SERS determined MIC value is the antibiotic concentration in red for the four bacterial strains that are susceptible to the indicated drug; *K. pneumoniae* 6908 and *E. faecium* 700221 are resistant to ampicillin and vancomycin respectively.

While peak and area SERS-AST profiles are essentially independent of drug concentration for resistant bacterial strains, it is perhaps not surprising that, in contrast to susceptible strains, the relative purine contributions to these SERS spectra are also independent of drug dose during incubation, e.g. *K. pneumoniae* 6908/ampicillin and *E. faecium* 700221/vancomycin (Figs. 7e and 7f). Presuming the metabolic origins of these SERS-AST signals, the dose independence of the molecular composition of these resistant strain SERS spectra confirms that the nucleotide degradation pathway is apparently not perturbed by antibiotics in resistant strains.

## IV. Discussion

The biochemical bases for this rapid SERS-AST technique are inherent bacterial responses that are intended to promote survival from two specific environmental stresses/threats; nutrient depravation and antibiotics. SERS purine signals result from secreted nucleotide degradation products initiated by starvation and may be viewed as the AST detection step. The dose-dependent re-wiring of bacterial metabolic responses during a 30-minute incubation with antibiotics prior to the onset of starvation enables the discovered AST capability of SERS resulting in the > 80% decrease in SERS intensity at the MIC. A more detailed biochemical description of the bacterial responses and other coincident chemical physics factors, that account for the success of this ultra-rapid SERS AST methodology are discussed below.

### A. Purines dominate the 785 nm bacterial SERS spectra

**1. Stringent response and purine secretion.** In order to survive under unfavorable conditions, bacteria have evolved with complex genetic networks, which allow them to sense and rapidly respond to environmental threats/stresses. We previously attributed the observed 785 nm SERS signals on Au substrates to the purine bases and some nucleosides, resulting from the catabolic degradation of nucleotides triggered by the ubiquitous bacterial stringent response (SR).[23-25, 37] This response is initiated when bacteria are placed in the no-nutrient environment of the water washing solutions during sample preparation (Fig. 1). The degradation of the so-called stable RNAs, *r*RNA and *t*RNA, which together account for ~96% of bacterial cell RNA, are known to be rapidly degraded upon nutrient starvation, most notably following amino acid starvation.[29, 39-43] This degradation begins as soon as nutrients become limiting, and notably, before bacterial cell growth ceases.[39] The intracellular nucleotide concentration available for further degradation thus quickly increases due to this SR induced RNA degradation. When the starved bacterial cells are subsequently placed on nanostructured SERS substrates, the SERS signal due to the secreted purine end products of this nucleic acid degradation process are correspondingly largest when scattering is collected from these near cell regions where their concentration is largest, at least initially.[30] Judging by the absolute size of the observed bacterial SERS signals, the local concentrations at the outer bacterial cell region are in the ~1 – 10 μM range thus allowing single bacterial cell level SERS signal observations.[30] This biochemical origin of the 785 nm bacterial SERS spectra was previously supported by isotopic labeling, cell and supernatant SERS comparisons, mass spec analysis, enzyme substrate dependencies, spectral component fitting analyses (e.g. Fig. 7) and analysis of gene knockout strain SERS spectra lacking specific purine degradation enzymes.[23]

However, since this earlier report[23] additional mechanistic details concerning the bacterial starvation response have been uncovered providing further understanding for the purine dominance of the bacterial SERS spectra. The molecular level trigger for the SR is known to be a buildup of uncharged deacylated *t*RNA in the ribosomal A-site resulting from the lack of amino acids.[27, 28, 44, 45] In turn, this uncharged condition activates the synthesis of guanosine tetra- and penta-phosphate, jointly referred to as (p)ppGpp.[43] These signaling nucleotides are known as alarmones due their central role in slowing cell growth and metabolism adaptation to enhance survival starvation conditions in response to this environmental stress and are conserved across bacterial species.[46, 47] These alarmones mediate reduce protein production, and inhibit or enhance enzymatic activity in several metabolic pathways, including the purine metabolism pathways critical to this SERS-based methodology.[48-51]

An overview of the relevant purine nucleotide degradation/salvation catabolic/metabolic pathway is shown for a representative *E. coli* strain (MG1655) in Fig. 8 and illustrates the effects of elevated levels of (p)ppGpp resulting from the stringent response. The main purine reactants and products appear in the black boxes, and relevant enzymes are explicitly indicated in this figure. This nucleotide degradation network is given by the KEGG database[52] but has been augmented by the addition of the recently characterized, widely conserved PpnN enzyme (pyrimidine/purine nucleotide 5'-monophosphate nucleosidase).[53, 54] As shown, PpnN facilitates the one-step cleavage of nucleoside monophosphates resulting in a free nucleobase (X = guanine, hypoxanthine, xanthine, adenine) and D-ribose-5'-phosphate (R5P):

$$XMP \rightarrow X + \text{D-ribose (5RP)}. \quad (1)$$

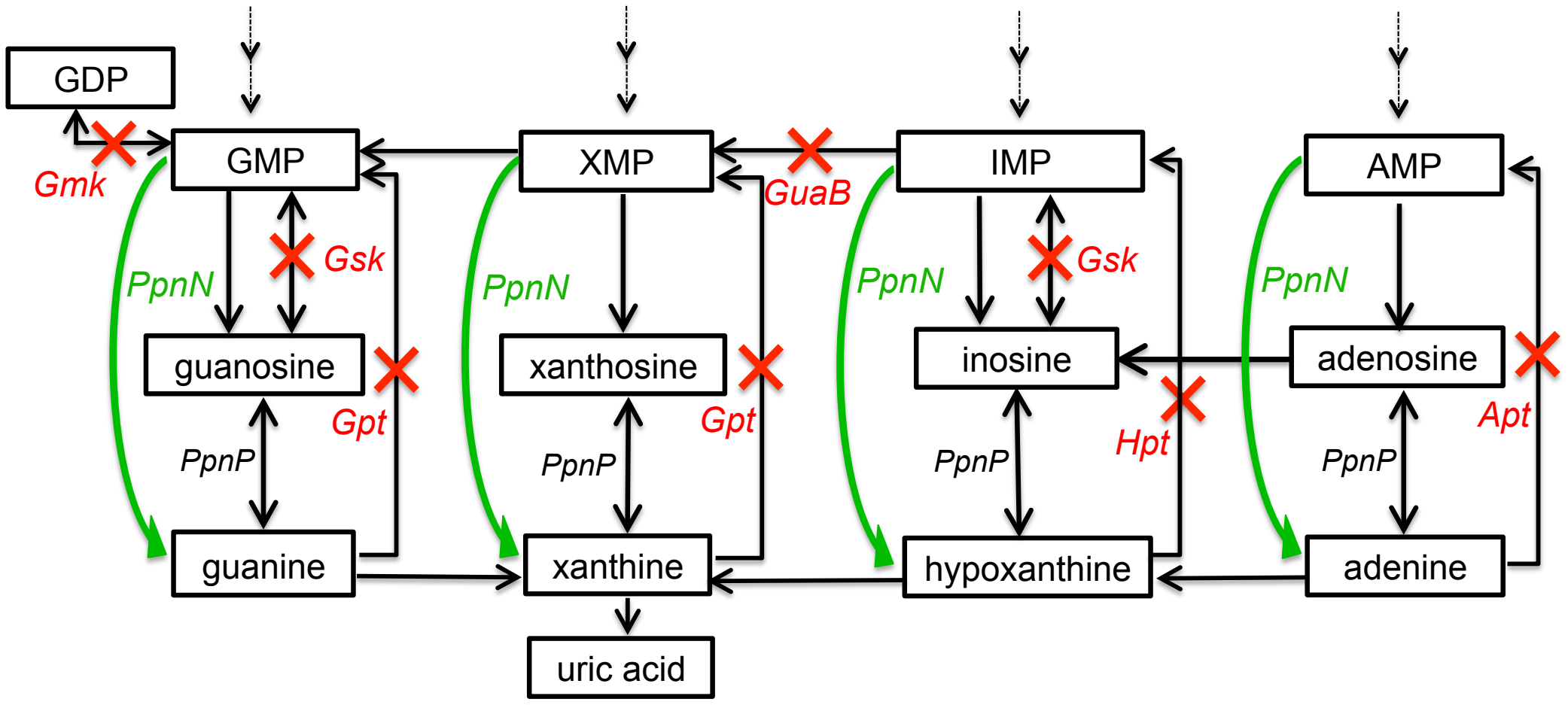

**Figure 8.** Summary of the purine metabolic/salvage mechanisms for a representative *E. coli* strain (MG1655) given by the KEGG pathway reference with the addition of the recently identified PpnN. Following RNA degradation, symbolized by dashed vertical arrows at top, all arrows correspond to enzymes facilitating the indicated reactions and equilibria between purinergic species. The green arrows and red crosses correspond to enzymes whose indicated transformations have been enhanced or inactivated respectively by (p)ppGPP over base activity levels. The net effect of this alarmone on this network is to funnel this RNA degradation towards the production of nucleobases and nucleosides that are secreted and observed by SERS.

Recent studies have revealed that (p)ppGpp strongly interacts with several proteins in this degradation/salvation network[55] that significantly impact the observed bacterial SERS signal production and resultant AST capabilities.[27, 29, 56] Following the SR onset, initiated when the bacteria are water washed, (p)ppGpp levels rise and bind to PpnN triggering a large conformational change exposing the protein's active site, thus accounting for a (p)ppGpp stimulatory effect on catalyzed nucleotide degradation[57] (green arrows in Fig. 8). Secondly, (p)ppGpp exerts an inhibitory effect (red X in Fig. 8) on the purine phosphoribosyltransferases (Gpt, Hpt, Apt) which are salvage pathway proteins that work in the opposite direction to PpnN converting nucleobases into nucleotides.[27, 29, 56-58] Additionally, (p)ppGpp also binds to several additional proteins (Gsk, Gmk, and GuaB) in this network inhibiting non-degradation activity as well (red X Fig. 8). The net effect of these SR initiated (p)ppGpp interactions is the rapid accumulation of free nucleobases. This alarmone enhances direct nucleotide degradation to nucleobases and blocks the return synthetic pathways for making new nucleotides from this excess (Fig. 8). Making nucleotides is energetically expensive[59] thus shutting down this purine biosynthetic pathway under low nutrient conditions enhances survival. Nucleobase secretion allows them to be taken up by bacterial cells and re-used in the one-step nucleotide salvage synthesis pathway (Gpt, Hpt, Apt) when nutrient rich environmental conditions return.[60]

Furthermore, it has recently been discovered that metabolism of R5P, the other product derived from (p)ppGpp enhanced PpnN activity (Eq. 1), results in increased amounts of key intermediates that lead to the production of aromatic amino acids within minutes of starvation triggering.[29] Thus, this nucleotide degradation product, R5P, can contribute to an important bacterial survival strategy under starvation stress conditions, with the ancillary formation of free nucleobases and the consequent SERS detection of purines secreted by viable bacterial cells.

Finally we note that equilibrium concentrations of nucleobases and their corresponding nucleosides (Fig. 8) are possible under starvation conditions since (p)ppGpp binding to the enzyme PpnP (Pyrimidine/purine nucleoside phosphorylase) is not observed and is consistent with the identified adenosine and guanosine contributions to some bacterial SERS spectra (Figs.

6, 7).[55] Thus, SR (p)ppGpp activity efficiently funnels the nucleotide degradation processes to result in the rapid accumulation of purine nucleobase end products, e.g. adenine, guanine, hypoxanthine, xanthine, uric acid, etc. which are then secreted and detected via SERS. This is a very rapid process; the majority of cellular *t*RNA and *r*RNA is degraded within twenty minutes following the onset of amino acid starvation[39, 40, 43] and the up-regulation of (p)ppGpp production and its consequent impact on the purine degradation network has been shown to occur on the timescale of 5 – 10 minutes.[47, 55, 57] These very fast timescales are consistent with other observations of rapid RNA depletion, (p)ppGpp dynamics and the prompt appearance of increased nucleobases in the bacterial metabolome.[17, 41, 48, 60-63] It's this rapid bacterial reprogramming response to starvation, in part, that sets the fast speed for this SERS-AST methodology.

**2. Additional factors contributing to purine dominated bacterial SERS signals.** Aside from the starvation induced biochemical mechanisms described above that account for the bacterial secretion of purines, other interactions also strongly contribute to the efficacy of this optical AST methodology. Mass spectrometry and HPLC reveal a wide variety of biological molecules in the extracellular region of bacterial cells,[23, 41, 60] and yet purines dominate the 785 nm bacterial SERS spectra. The small size and multiple lone pair of nitrogen electrons result in the very large ~785 nm excited SERS cross-sections of purines on both Au and Ag nanostructured surfaces relative to other metabolites such as amino acids, proteins, etc., present in the bacterial extracellular region.[23, 64-67] As mentioned above, only a small cytosine contribution could be identified in the SERS spectra of a few bacterial species, although uracil and cytosine have been reported in the metabolome of starved bacterial cells.[41] This spectral absence in part results from the much smaller SERS cross-sections of pyrimidines relative to purines (SI Fig. S16). The per molecule 785 nm SERS intensity of cytosine, uracil and thymine are respectively ~35, 175 and 150 times smaller than that of adenine, and thus these pyrimidines are difficult to detect relative to purine components. The single ring and fewer nitrogens may result in a weaker physi-adsorption or less favorable orientation with respect to the Au or Ag substrate surface thus accounting for the smaller SERS enhancement of pyrimidines relative to purines. Other biochemical factors may also play a role in the absence of pyrimidine base contributions to these SERS spectra given the large biological role of purinergic signaling and the importance of purine biosynthesis to bacterial stress responses.[38]

Secondly, the (p)ppGpp inhibited purine salvage phosphoribosyltransferases, Hpt, Gpt, Apt, (Fig. 8) are reported to be located at the cell membrane.[63] In addition to facilitating the X →XMP conversion, these enzymes are also thought to transport these reactants and products across the cell membrane.[63, 68, 69] Such cell wall localization may also contribute to the efficient and prompt appearance of secreted nucleobases in SERS spectra. These factors combine to make SERS exquisitely sensitive to the catabolic degradation of the tightly regulated intracellular concentrations of nucleotides resulting from the bacterial SR, and significantly account for the nearly exclusive dominance of purine nucleobases contributing to the 785 nm SERS spectra of bacteria fortuitously allowing their direct use for providing rapid quantitative MICs.

**B. Antibiotic perturbations alter secreted purine concentrations.**

**1. Effects of antibiotics: General considerations**. The above discussion only provides a mechanistic basis for the bacterial responses, mediated by the SR (p)ppGpp alarmone, resulting in the rapid secretion of purines upon starvation and thus the observed 785 nm bacterial SERS spectra. In contrast, the biochemical details explaining how the relative magnitude of these responses are affected by a *pre*-starvation 30-minute exposure to antibiotics, which is the basis of the quantitative SERS-AST methodology (Figs. 2 – 5), is not fully established. However, results reported here and prior studies offer insights regarding the relevant mechanisms that can be controlling the observed dose dependence of bacterial secreted purines, i.e. the SERS intensities, when starvation is initiated *after* antibiotic exposure.

We first note that since bacterial SERS signals for susceptible strains uniformly exhibit decreased intensities (>80%) at the ~MIC (Figs. 2– 5) across the diverse classes of antibiotics, the mechanism for this SERS effect cannot be due to the different primary inhibitory initial targets for these specific antibiotics. Furthermore, given this SERS intensity decrease at the MIC is observed for all tested bacterial samples (Figs. 2 – 5),[31, 32] this response to antibiotics appears to be highly conserved allowing it to be exploited for development of a clinically relevant rapid AST platform. Recent studies have established that antibiotic treatments significantly alter the metabolic state of bacteria..[38, 59, 70-74] and importantly, there is an emerging recognition that dose dependent metabolic perturbations far removed from their primary interaction target play a very large role in antibiotic efficacy and bacterial death processes.[38, 59, 71-74] In particular, nucleotide metabolism appears to be a well-conserved mechanism that bacteria have evolved to respond to diverse stresses such as starvation, discussed above, as well as antibiotics.[75] Furthermore, just as found for the SERS response to antibiotics, the diverse range of bactericidal antibiotics induce similar metabolic changes.[71] Thus we attribute the dose dependent effects of antibiotics on the purine SERS signals triggered by the starvation response, to such downstream secondary processes, i.e. altered purine metabolic pathways, in response to antibiotic exposure during the 30-minute incubation period with these drugs.

**2. Evidence that nucleotide degradation is altered by antibiotics.** Prior evidence already indicates that antibiotics perturb the nucleotide biosynthesis/degradation pathways.[71, 76] For example, rapidly reduced *intracellular* nucleotide, nucleoside and purine nucleobase pools (e.g. adenine and guanine) have been reported following antibiotic treatment[38, 71, 73] Such decreased concentrations, were reported in *E. coli* at 30 min after drug exposure for a range of antibiotics with different modes of initial interaction, consistent with the timescales for this novel AST methodology based on secreted purine levels. Nucleotide metabolic intermediates are among the most recurrently affected metabolites following exposure to a wide range of antibiotics.[76] Correspondingly, proteomics studies have found proteins involved in most metabolic pathways, including purine metabolism, were down regulated in response to specific antibiotics (tetracycline) and RNA transcription levels, the hypothesized sources of the observed secreted purines, have also been reported to be rapidly down regulated in response to antibiotics in specific studied systems.[17, 77]

Additionally, it has been shown that antibiotics can alter (p)ppGpp levels, the key SR alarmone that in part controls the efficiency of the nucleotide degradation pathway (Fig. 8) as discussed above. Some antibiotics have been shown to block the rise of (p)ppGpp preventing ribosomal RelA activation of the SR response[77-80] while manipulation of (p)ppGpp levels has been shown to convert chloramphenicol from a bacteriostatic to a bactericidal drug. These observations are consistent with the secondary effects of antibiotics on nucleotide metabolic pathways resulting in dose dependent changes to the levels of secreted purine nucleobases, the basis for this AST methodology.

**3. Bacteriostatic and Bactericidal specific considerations.** Part of the challenge for understanding how SERS intensities result in quantitative MIC values following a 30 min antibiotic exposure period is that not all the mechanistic details of antibiotic induced cell death/stasis are fully determined. However, some established biochemical mechanisms of bacterial responses to antibiotics support the role that secondary metabolic pathways disruptions play in the SERS effects reported here. For example, bacteriostatic antibiotics have been shown to suppress bacterial cellular respiration, whereas most bactericidals result in accelerated respiration.[73] We find this observed bactericidal/bacteriostatic increase/decrease in bacterial respiration rates correlates with the observed SERS intensity changes at sub-MIC dosages. As seen in Figs 2 – 5, the SERS detected purine secretion levels are enhanced relative to no drug exposure at sub-inhibitory doses for bactericidals and mostly monotonically decrease for bacteriostatics (Figs. 2 – 5). Furthermore, these antibiotic induced respiration rate changes are fast (~tens of minutes), consistent with these SERS effects.[73]

Reported decelerated bacterial respiration rates in response to bacteriostatics are dose dependent and maximally achieved at the MIC concentration with no substantial changes at higher concentrations[73] mimicking the dose dependent SERS intensity trends observed for bacteriostatic exposure (Figs. 2 – 5). The predominant cellular process initially targeted by bacteriostatics is translation resulting in reduced protein synthesis.[59, 73] The observed bacteriostatic cellular respiration reduction may be a byproduct, in part, of this translation inhibition. In terms of the nucleotide degradation pathways, drug-induced inhibition of translation during the 30 minute incubation decreases amino acid consumption, which can increase *t*RNA aminoacylation levels (charged RNA) during this exposure period.[71] A resulting potential consequence is that the RelA response,[81] triggered by the accumulation of uncharged *t*RNA, is thus muted as starvation conditions are encountered after the drug incubation period in the SERS-AST procedure. Correspondingly (p)ppGpp alarmone levels would be reduced relative to no bacteriostatic exposure and hence dose dependent secreted purine levels would be diminished consistent with the SERS results reported here (Figs. 2 – 5).[44] In further support of this mechanism, amino acid accumulation is observed in the metabolome of bacteriostatic exposed bacteria.[73] The excess amino acid pool might also obviate the need for 5RP resulting from the (p)ppGpp enhanced degradation of nucleotides (Eq. 1), possibly also contributing to reduced nucleobase secretion levels after bacteriostatic exposure. As discussed 5RP is a potential precursor to *de novo* amino acid synthetic pathways.[29] Thus, dose dependent translation inhibition leading to decreased consumption and build up of intracellular amino acids, can *reduce* the alarmone concentration, with the resulting effect of dose dependent reduced SERS intensities immediately following placement in water.

In contrast, measured respiration rates are accelerated at low doses and are a maximum at the MIC for bactericidals. Both increases and decreases in relative metabolic activity of different pathways are reported in response to bactericidals, indicating that they result in more complex perturbations of metabolism as the SERS AST profiles analogously reflect.[71] The demands of *accelerated* respiration and the corresponding *enhanced* cellular metabolic rates of some processes induced by bactericidals appear to result in the SERS detected higher levels of secreted purines at sub-inhibitory concentrations (Figs. 3- 5). The abundance of intracellular central carbon metabolites and the disruption of the nucleotide pool said to be consistent with accelerated nucleotide turnover rates are reported metabolic consequences of bactericidal antibiotics after 30 min.[71, 76, 82] This clearly parallels trends in the SERS-AST profiles for bactericidals. Furthermore, the up-regulated demand for central carbon metabolism, correlated with higher respiration rates in response to bactericidals, can also benefit from the extra amino acid precursor, 5RP, resulting from nucleotide degradation (Eq. 1) thus additionally contributing to increased nucleobase concentrations for secretion. This sub-inhibitory response to bactericidals, i.e. increased respiration and corresponding metabolic rates, has been hypothesized to be an adaptive mechanism for surviving low levels of naturally occurring antibiotics.[72] However, at least two metabolic mechanisms act must act with opposite effects for bactericidals since SERS purine intensities plunge at the MIC after increasing at sub-inhibitory concentrations. At higher antibiotic concentrations near the MIC other secondary antibiotic metabolic perturbations are found to contribute to lethality through mechanisms that involve the generation of reactive oxygen species (ROS) and other harmful molecules.[38, 59, 71, 73, 74] These reactive species damage many important cellular components, but oxidation of the nucleotide pool appears to be particularly significant,[83] and thus reduced nucleobases and nucleosides concentrations may be available for secretion due to direct and indirect cellular damage from ROS. While the rapid depletion of free *intra*cellular nucleobases (e.g. adenine, guanine, and cytosine) in response to bactericidals.[38, 71] has been noted above, adenine depletion in particular is thought to increase ATP demand via purine biosynthesis, resulting in elevated central carbon metabolism activity and oxygen consumption, thus enhancing the killing effects of these antibiotics. Adenine is a nearly ubiquitous component of 785 nm SERS spectra. Hence the decrease in secreted purine levels evident in the SERS-AST profiles may not just be an ancillary effect of global metabolic slowdown in response to antibiotic exposure but symptomatic of some key metabolic responses that contribute to bacterial cell death. ROS may be involved in damage to (p)ppGpp, which would also contribute to decreased purine secretion levels at higher bactericidal concentrations.[72] The SERS-AST profiles reflect this complex dose dependent relationship between the metabolic rewiring in response to bactericidals, and the purine biosynthetic and alarmone (p)ppGpp mediated degradation pathways.

## V. Conclusion

To our knowledge this SERS based methodology incorporating a 30-minute incubation period with antibiotics in growth media, water washing and signal acquisition on Au nanoparticle substrates is *the fastest demonstrated phenotypic scheme for accurate MIC determinations* and readily distinguishes drug susceptible and resistant strains. Quantitative AST profiles result in ~1 hour for both Gram positive and negative species, and all classes of initial drug-bacterium targets. More specifically SERS-AST results from the rapid rewiring of metabolic pathways, as bacteria respond to specific antibiotics during a 30-minute incubation period, altering the consequences of the stringent response once starvation is subsequently triggered thus affecting secreted purine levels of degraded RNA. This downstream secondary nucleotide degradation response is unperturbed in resistant bacterial strains. The fundamental bases for this success is due to several fortuitously coincident phenomena: (1) the stringent response is rapid and ubiquitous across all bacterial species, (2) purine bases are very bright markers that dominate the 785 nm SERS spectrum, and (3) the complex bacterial metabolic response to antibiotics is both rapid, dose-dependent and tightly coupled to nucleotide degradation pathways.

Given the previously demonstrated ability to observe SERS spectra of single bacterial cells[30] and bacteria enriched from infected human blood at clinical levels (2 – 11 cfu/mL) after 5 hours of growth,[84] we anticipate that the parallelized and automated embodiment of the ~1 hour SERS-AST approach will enable quantitative (MICs), phenotypic AST for blood infection treatments in < 8 hours, instead of ~2 – 3 days. When every hour delay leads to higher mortality rates, such as for blood stream infections,[7] this SERS based methodology has the potential to be transformative in treating this class of bacterial infections. Similarly, SERS-AST may be developed for UTI, STD or wound treatments filling a currently unmet need for ultra-rapid (~1 hour), phenotypic, antibiotic drug susceptibility determinations for other bacterial infections presentations. Its widespread adoption could result in more effective treatments, fewer chronic infections, reduced healthcare costs and importantly reduced AMR proliferation.

Leveraging the rapid, SERS reporting on purine degradation and biosynthetic pathways can provide a novel molecularly specific probe to further our understanding of bacterial antibiotic responses which is essential for developing new infection treatment approaches and for learning about the onset and survival mechanisms of bacterial persister cells[85, 86] in drug treated bacterial populations. For example, SERS at the already demonstrated single cell level[30] has the ability to evaluate susceptibility/resistance homogeneity within a given colony as a function of such variables as antibiotic exposure or starvation times, or other environmental changes with, at least, purine molecular specificity. The alarmone (p)ppGpp, which plays such a key role in controlling SERS purine signal intensities in response to starvation and antibiotics as described here, has also been found to play a central role in bacterial persistence development.[46, 75, 86-88]

As mentioned above, one other group has demonstrated the use of reduced SERS signal intensities to achieve AST in part affirming the robustness of these antibiotic sensitive results.[31, 32] However, higher bacterial concentrations on Ag substrates following ~2 hour incubation period are employed in those prior studies distinguishing the performance of these two procedures. Aside from the molecular and biochemical origins of the SERS-AST methodology offered here, the shorter time scale and higher sensitivity of the Au nanoparticle based methodology described in this report offers significant practical advantages for the best implementation of this SERS-AST approach.

**Declaration of Interests**

LDZ is scientific co-founder of Forecast Diagnostics Inc., an AST startup company.

**Author Contributions**

Conceptualization: LDZ; Data acquisition: AF; Data analysis: LDZ; Writing: LDZ

**Acknowledgement**

Support of a Boston University Ignition Award and a Boston University Clinical & Translation Science Institute Award is gratefully acknowledged.

**Table I. Antibiotics used in this study**

| Antibiotic | Class (subclass) | Bactericidal/ Bacteriostatic | Target of action |
|---|---|---|---|
| Levofloxacin | Quinolone | Bactericidal | DNA replication |
| Tetracycline | Tetracycline | Bacteriostatic | protein synthesis |
| Ampicillin | β-Lactam(Penicillin) | Bactericidal | cell wall synthesis |
| Vancomycin | Glycopeptide | Bactericidal | cell wall synthesis |
| Nitrofurantoin | Nitrofuran | Bacteriostatic | citric acid cycle & DNA, RNA, and protein synthesis |
| Tobramycin | Aminoglycoside | Bactericidal | protein synthesis |
| Chloramphenicol | Chloramphenicol | Bacteriostatic | protein synthesis |
| Clindamycin | Lincosamides | Bacteriostatic | protein synthesis |
| Meropenem | β-Lactam(Carbapenem) | Bactericidal | cell wall synthesis |
| Ciprofloxacin | β-Lactam(Cephalosporin) | Bactericidal | DNA gyrase inhibitor |
| Sulfamethoxazole /trimethoprim | Sulfonamide/ dihydropyrimidine | Bactericidal | Folic acid metabolism |

# Supplementary Information

# Ultra-rapid, Quantitative Antibiotic Susceptibility Testing via Optically Detected Purine Metabolites

A. Fraiman and L. D. Ziegler*

Department of Chemistry and The Photonics Center, Boston University, Boston MA 02215

* Corresponding author: lziegler@bu.edu

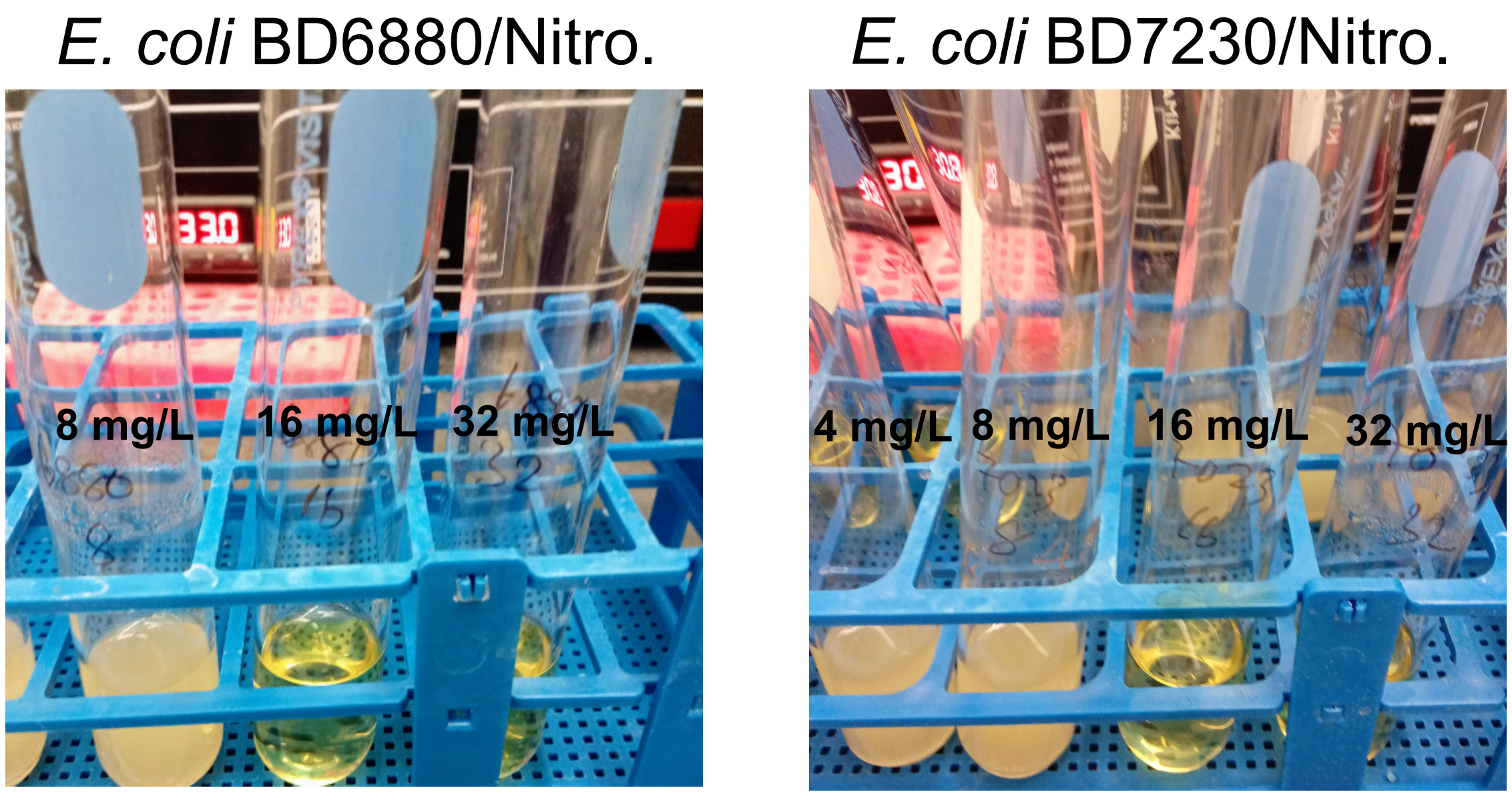

Fig. S1. Examples of MIC determination via CLSI 24 hour growth broth dilution procedure and visual inspection for *E. coli* 6880/nitrofuratoin and *E. coli* 7023/nitrofuratoin. The MIC corresponds to the lowest antibiotic concentration where turbidity in the growth media plus antibiotic solution sample vanishes. Thus, as shown in these pictures the MIC is16 mg/L for both *E. coli* 6880 and *E. coli* 7230 in nitrofurantoin.

Figures S2 – S14 are the SERS spectra of the bacterial strain/antibiotic pairs corresponding to the bar graphs shown in Figs. 4 and 5. These spectra have been offset by one unit to better show the changes in relative intensity as a function of antibiotic dose. These spectra were obtained following the same procedure used to acquire the SERS spectra in Figs. 2 and 3, and outlined in Fig. 1 and described in the *Materials and Methods* section.

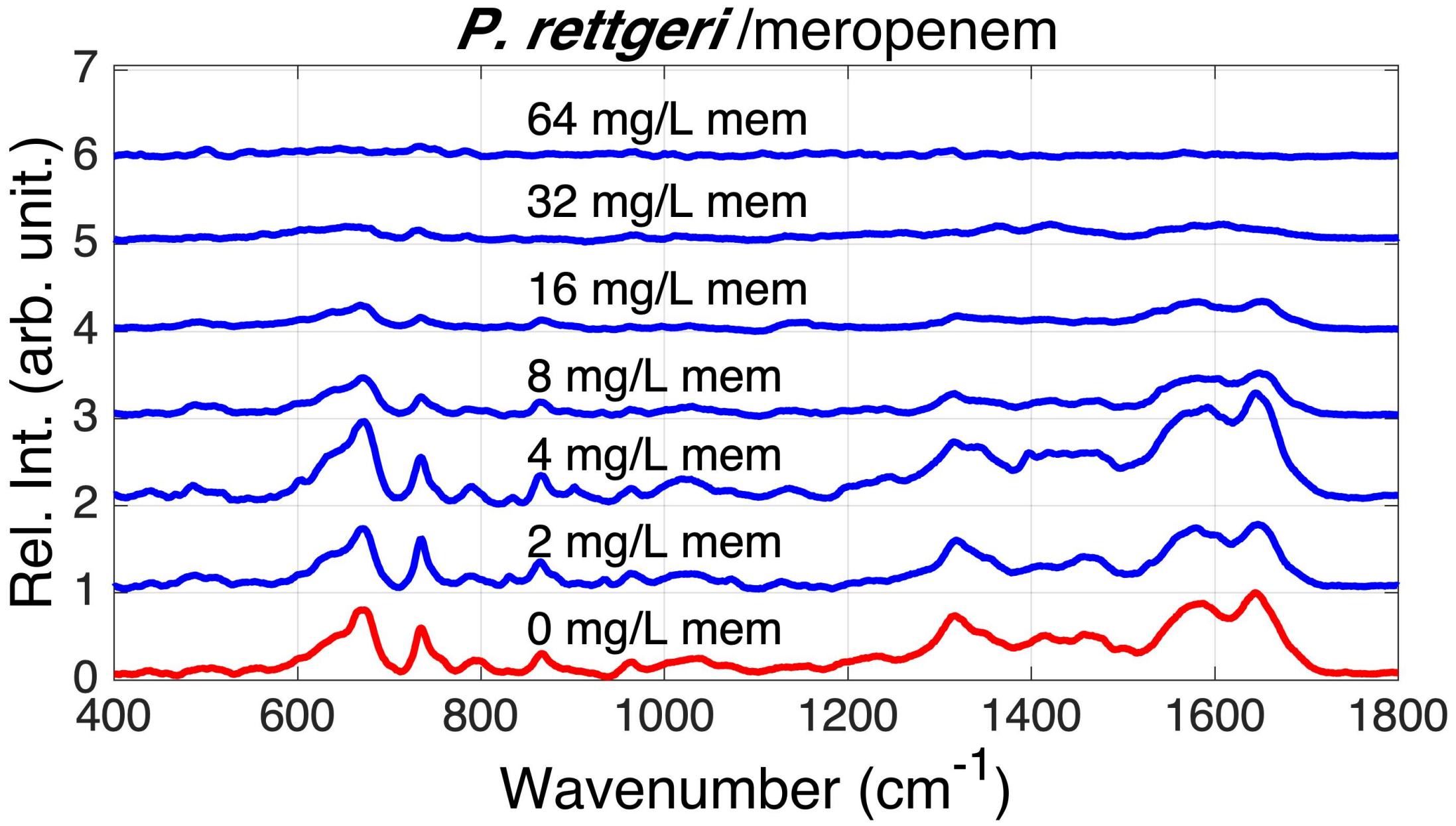

Fig. S2. SERS spectra of *P. rettgeri* 2525 as a function of meropenem concentration in growth media during 30 minute incubation.

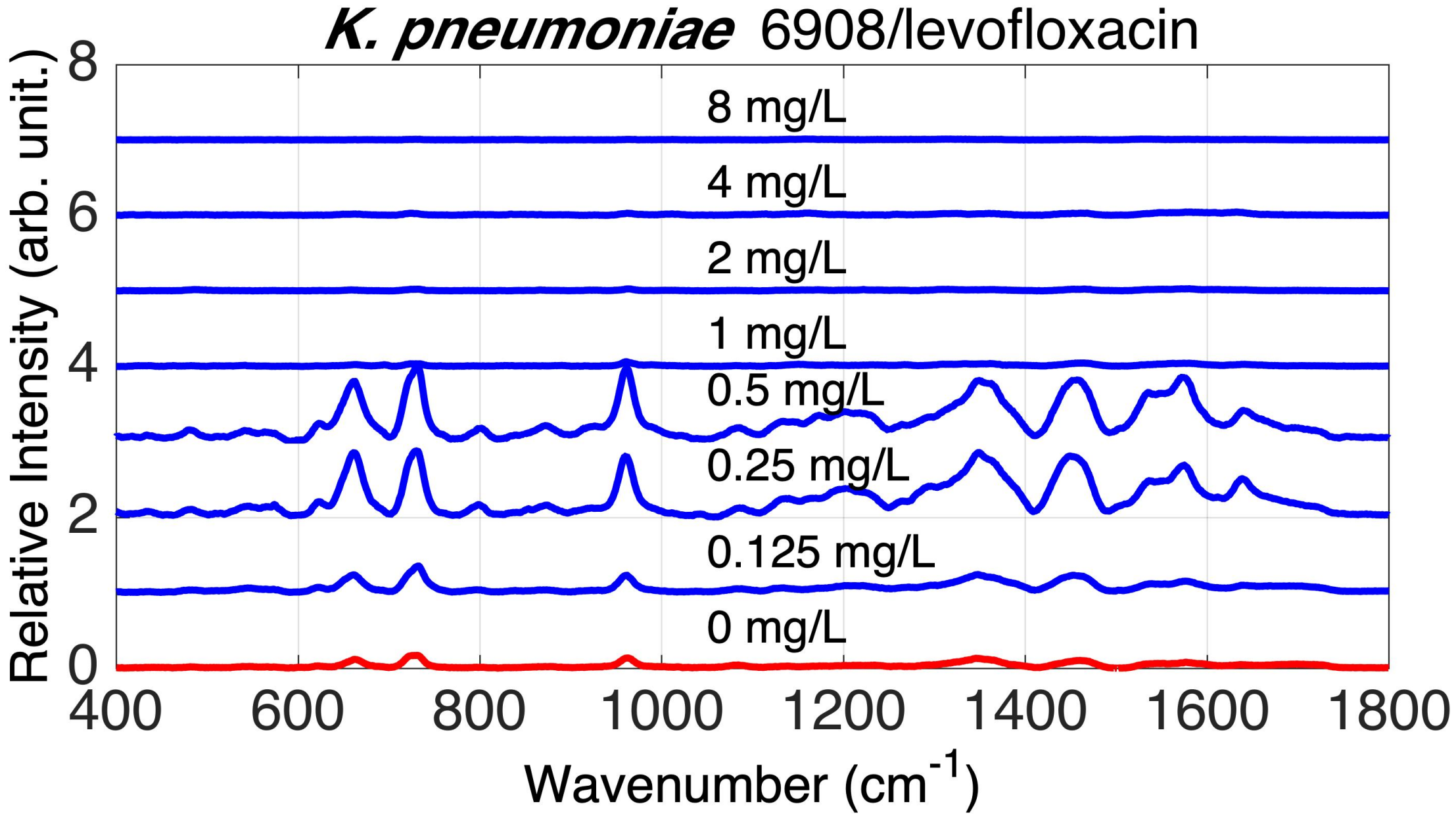

Fig. S3. SERS spectra of *K. pneumoniae* 6908 as a function of levofloxacin concentration in growth media during 30 minute incubation.

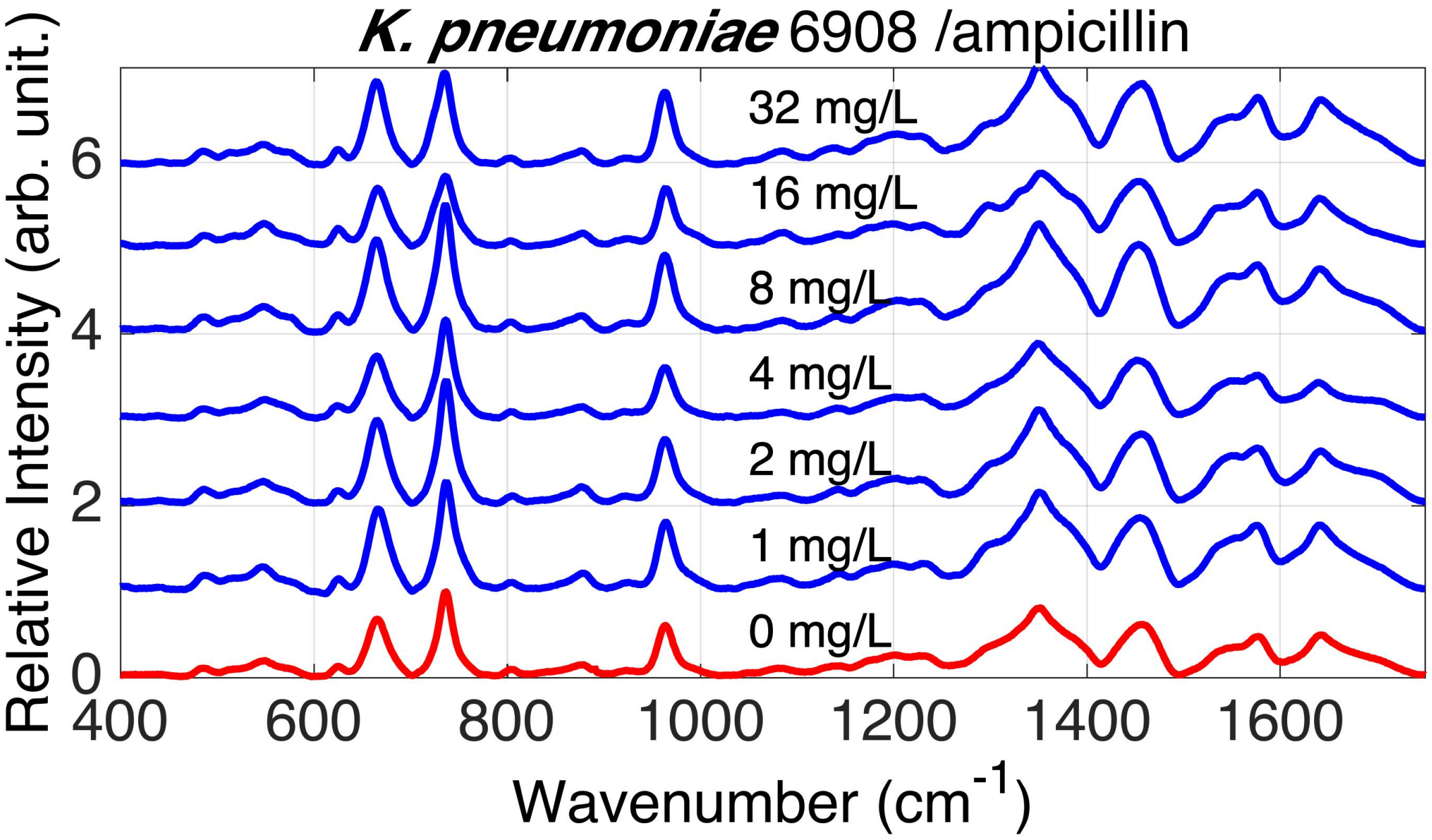

Fig. S4. SERS spectra of *K. pneumonia* 6908 as a function of ampicillin concentration in growth media during 30 minute incubation.

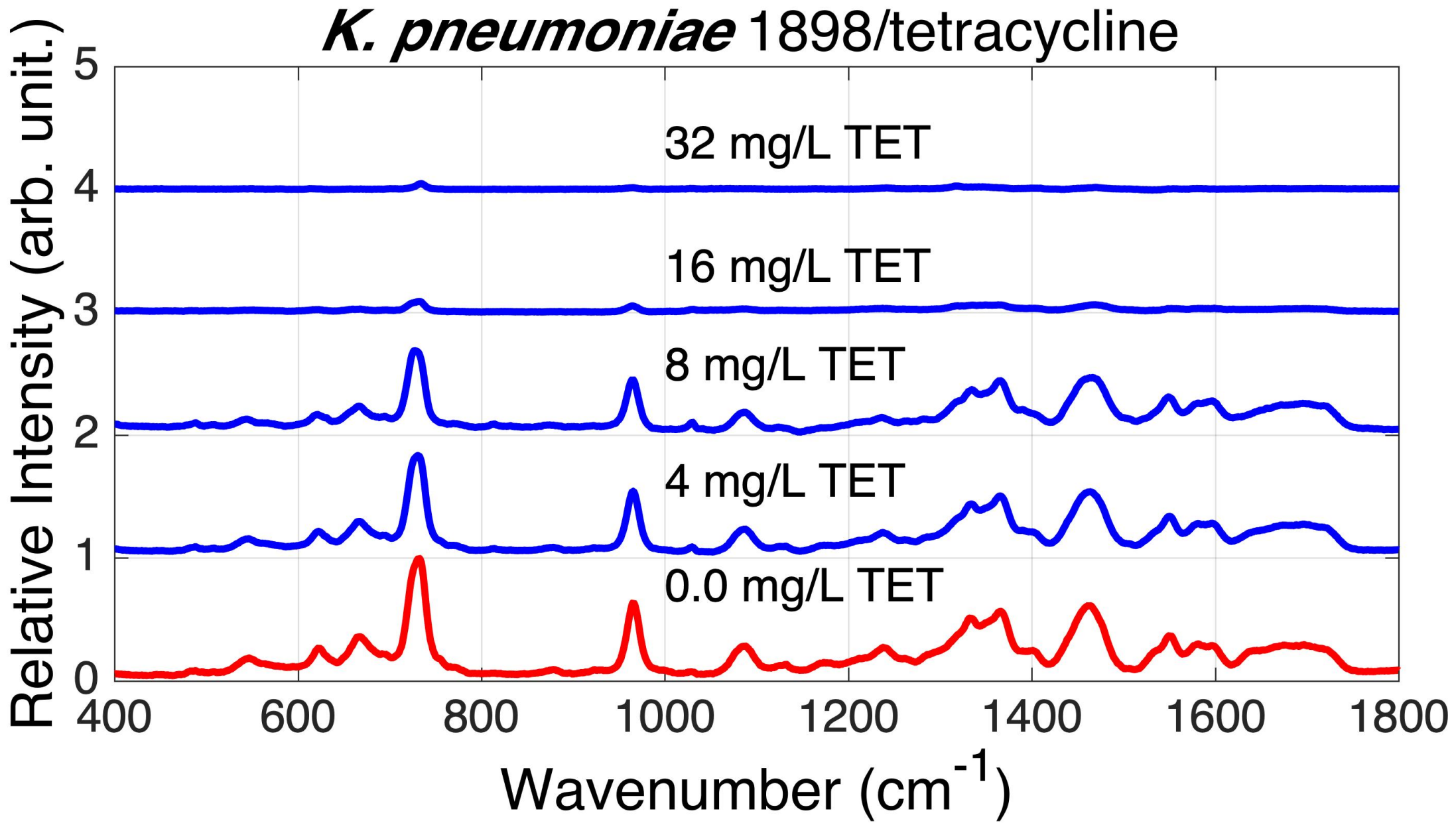

Fig. S5. SERS spectra of *K. pneumoniae* 1898 as a function of tetracycline concentration in growth media during 30 minute incubation.

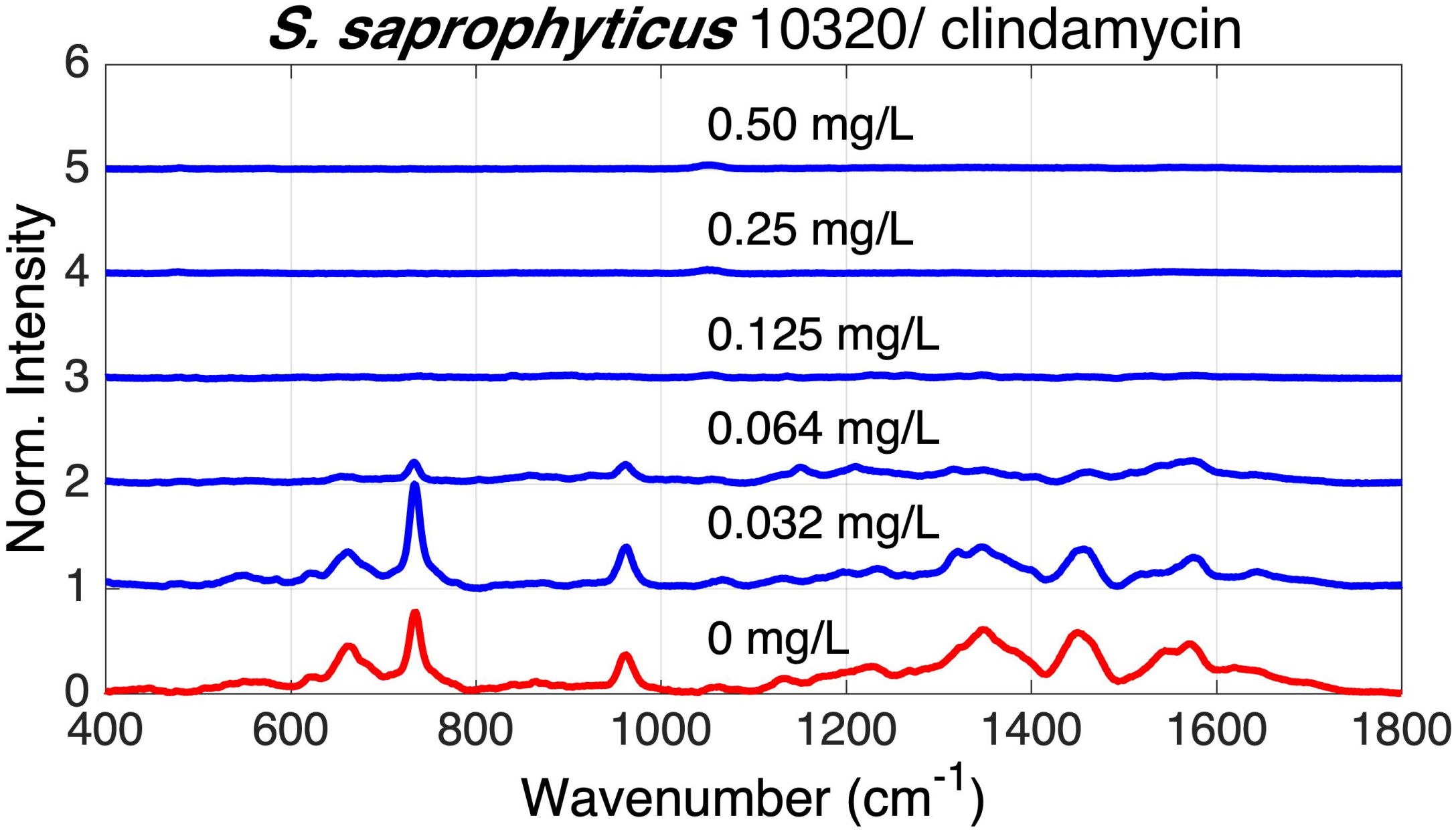

Fig. S6. SERS spectra of *S. saprophyticus* 10320 as a function of clindamycin concentration in growth media during 30 minute incubation.

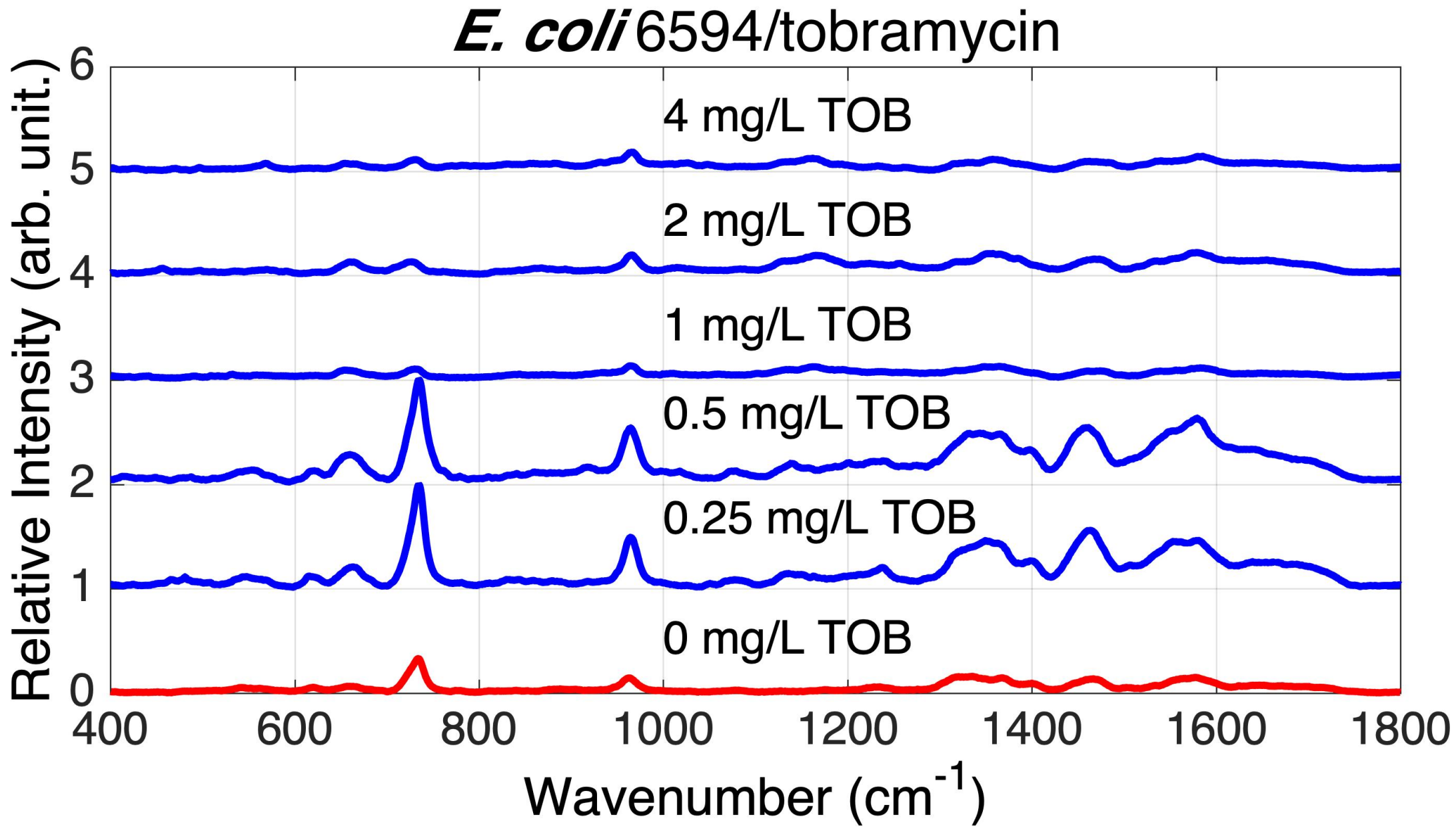

Fig. S7. SERS spectra of *E. coli* 6594 as a function of tobramycin concentration in growth media during 30 minute incubation.

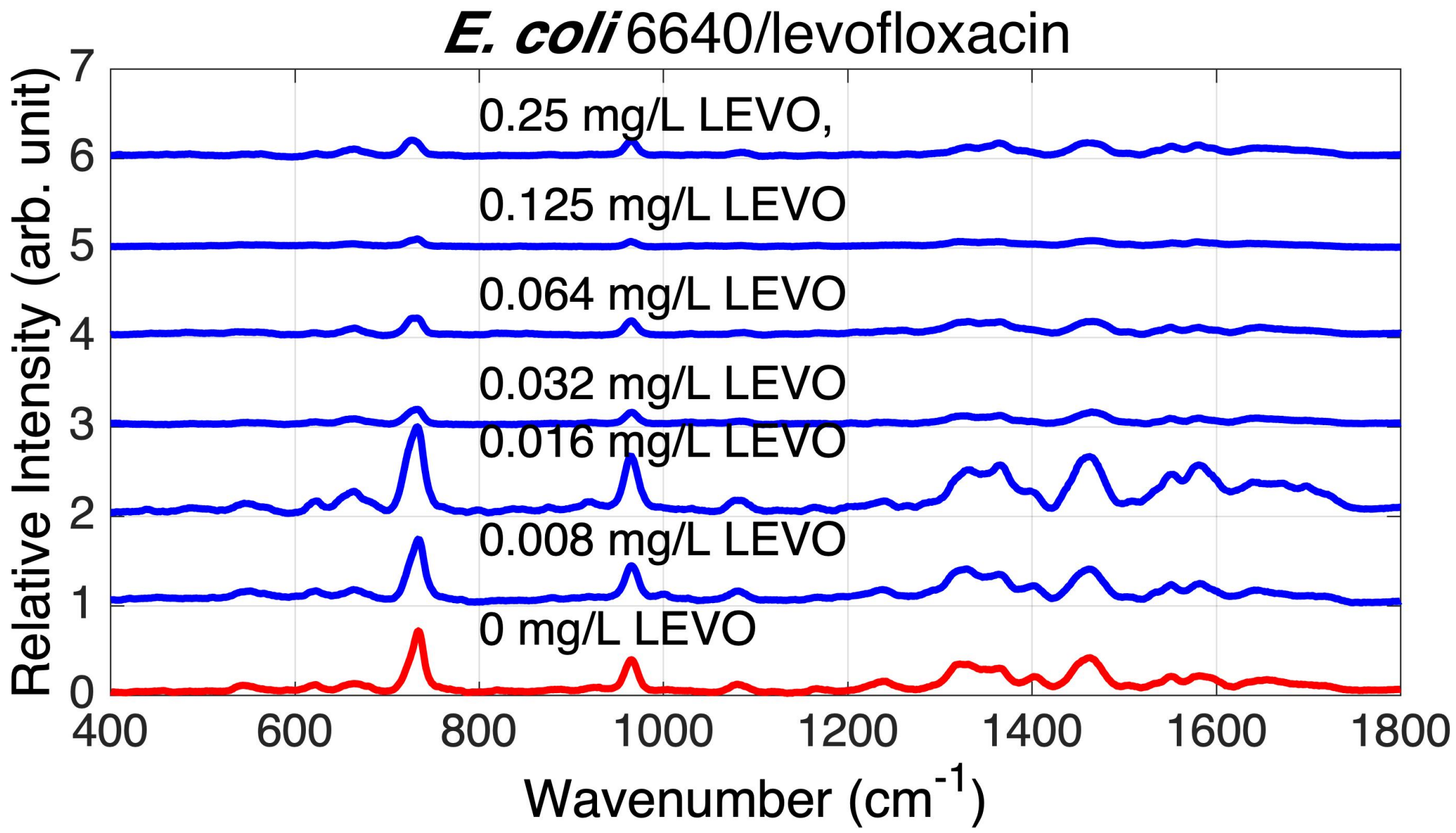

Fig. S8. SERS spectra of *E. coli* 6640 as a function of levofloxacin concentration in growth media during 30 minute incubation.

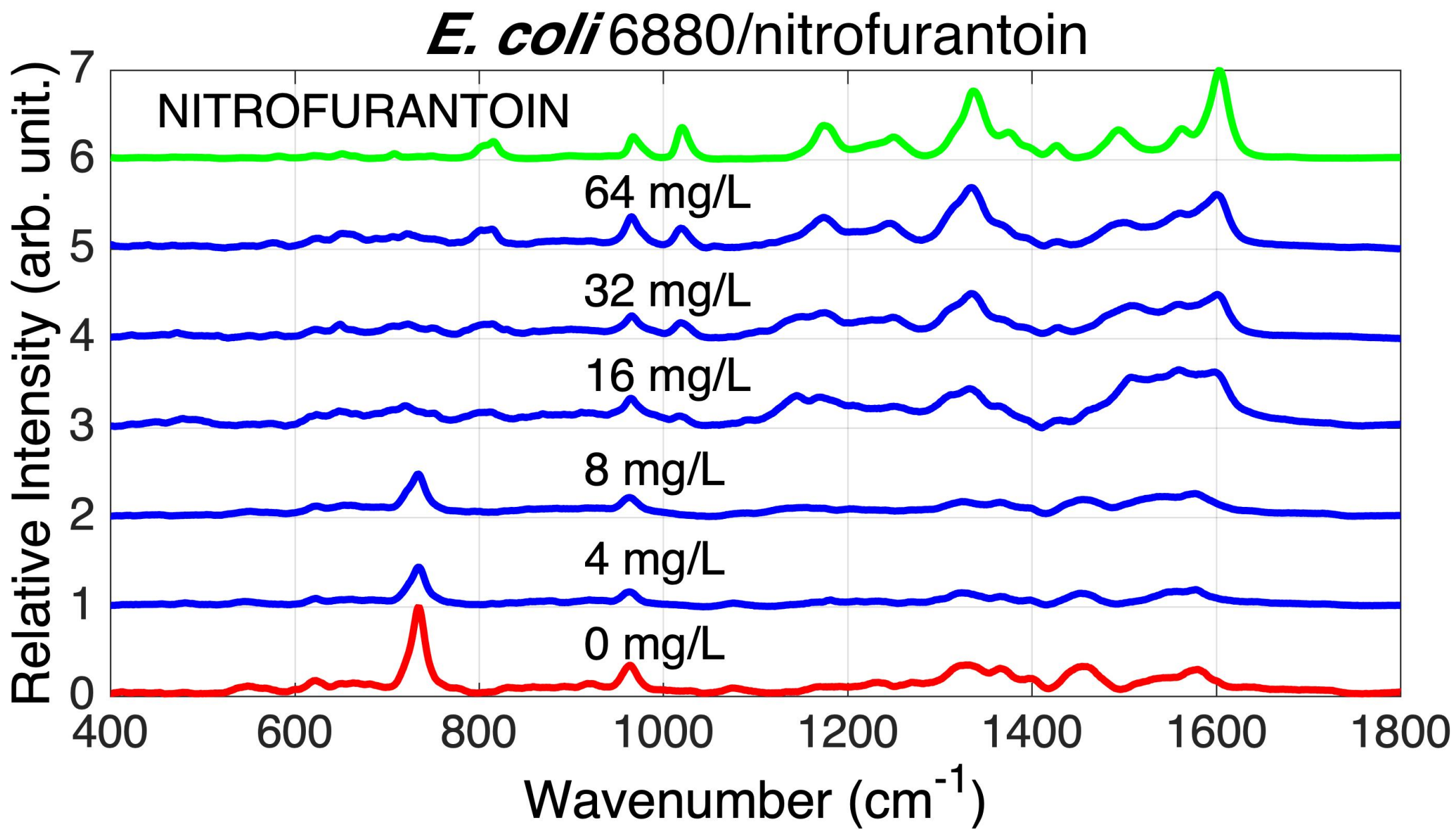

Fig. S9. SERS spectra of *E. coli* 6880 as a function of nitrofurantoin concentration in growth media during 30 minute incubation. Contributions of nitrofurantoin evident in higher concentration SERS spectra.

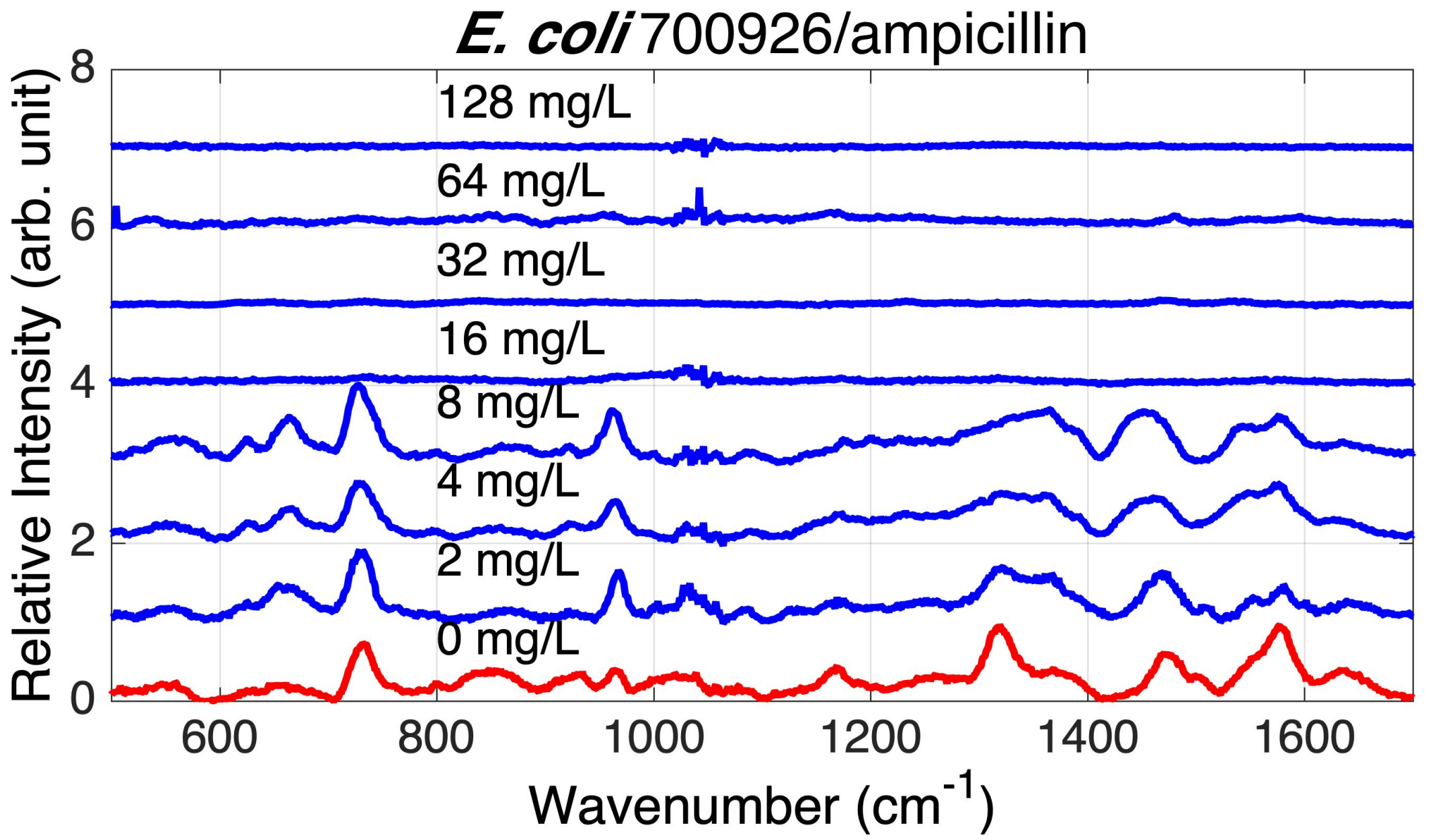

Fig. S10. SERS spectra of *E. coli* 700926 as a function of ampicillin concentration in growth media during 30 minute incubation.

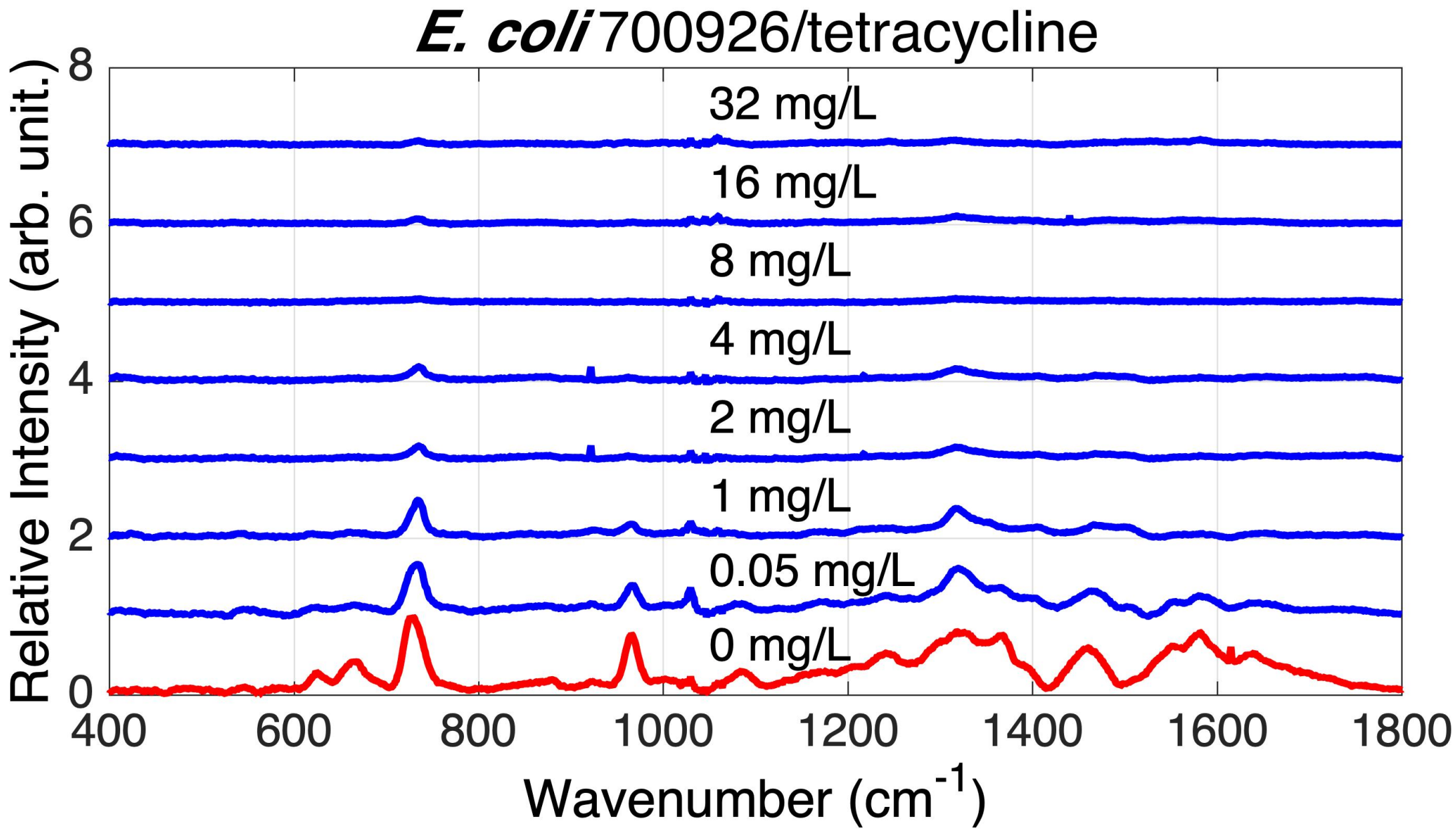

Fig. S11. SERS spectra of *E. coli* 700926 as a function of tetracycline concentration in growth media during 30 minute incubation.

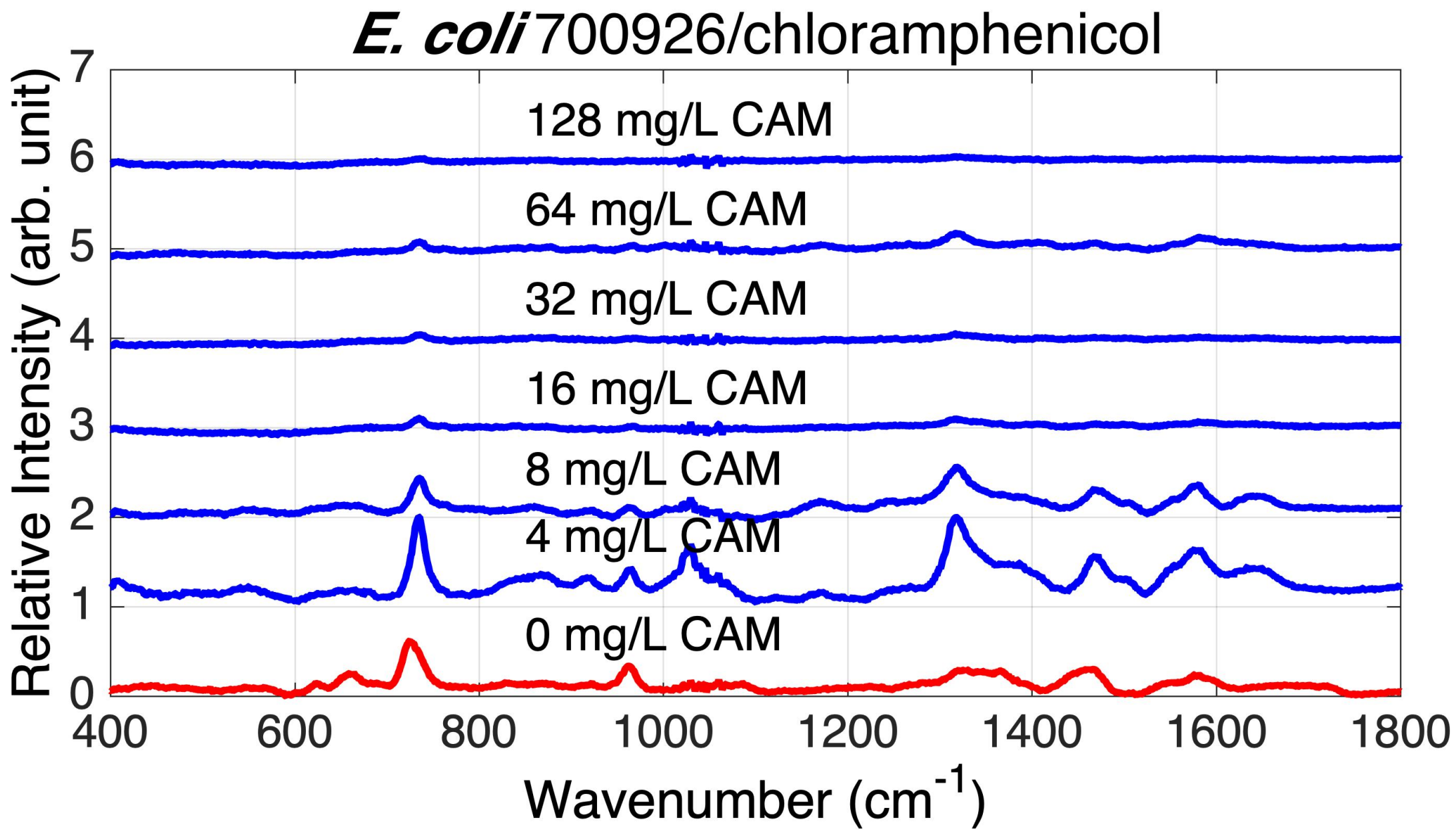

Fig. S12. SERS spectra of *E. coli* 700926 as a function of chloramphenical concentration in growth media during 30 minute incubation.

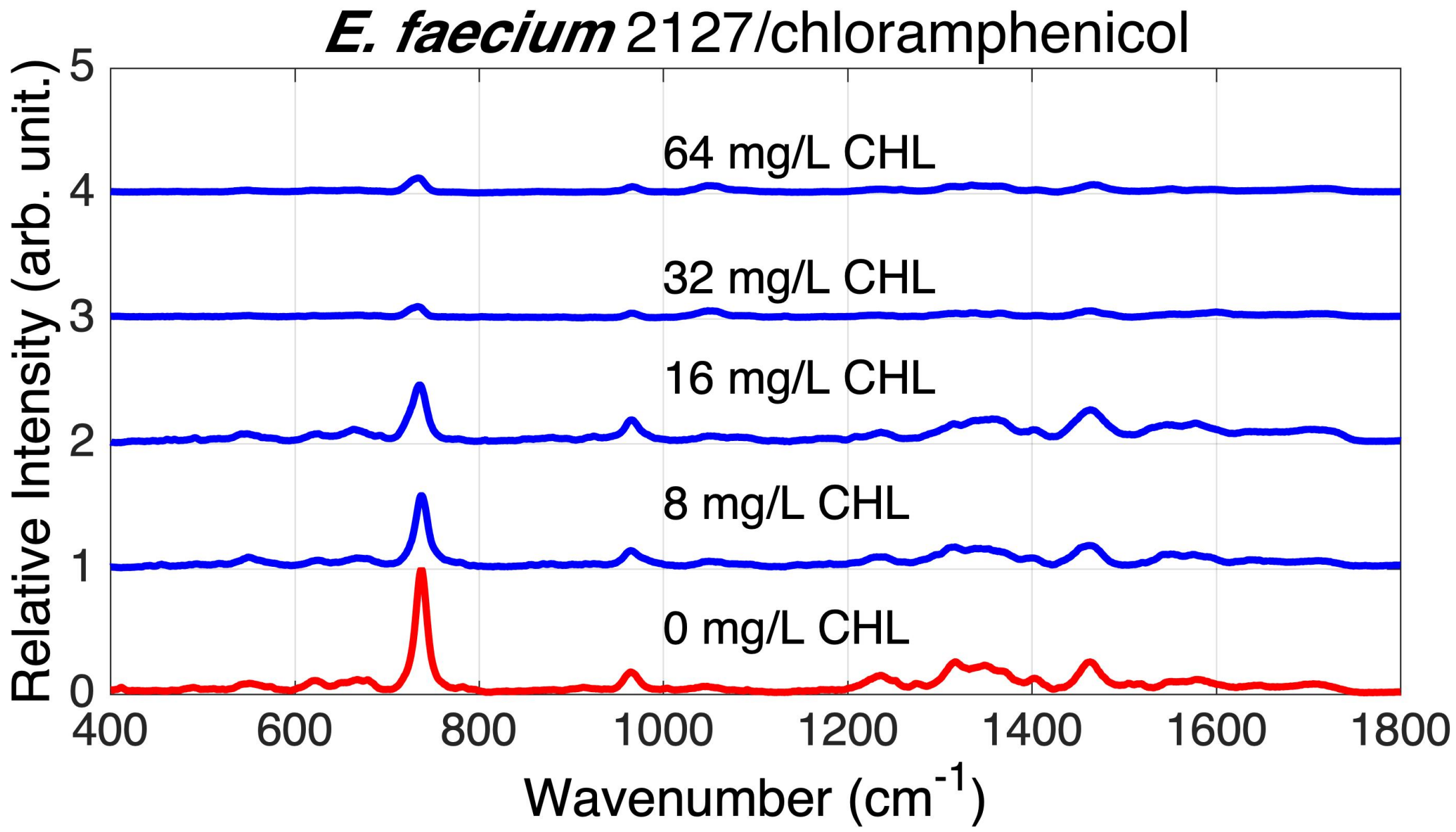

Fig. S13. SERS spectra of *E. faecium* 2127 as a function of chloramphenical concentration in growth media during 30 minute incubation.

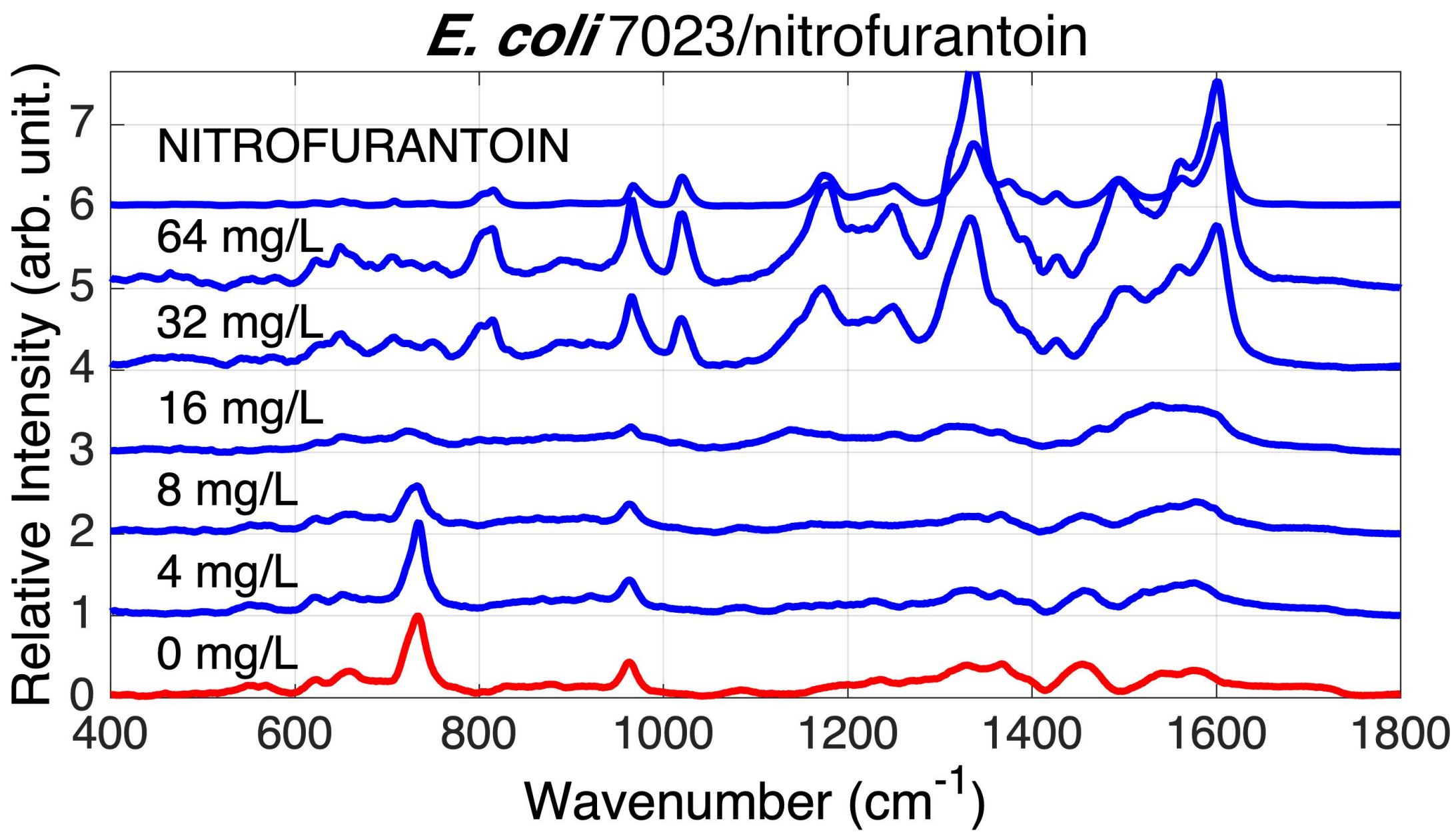

Fig. S14. SERS spectra of *E. coli* 7023 as a function of nitrofurantoin concentration in growth media during 30 minute incubation.

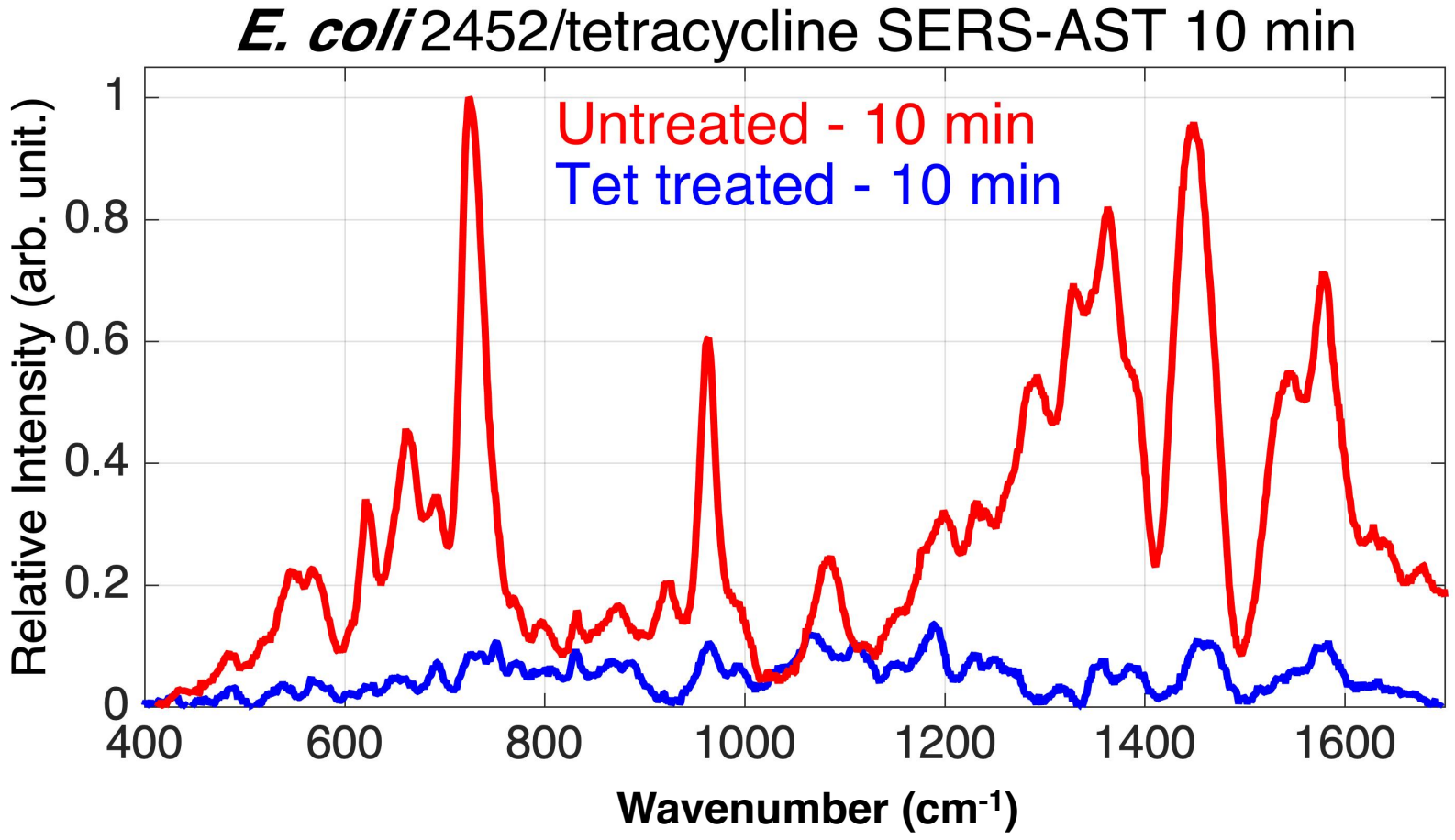

Fig. S15. SERS spectra of *E. coli* 2452 after 10 minute exposure to tetracycline compared to simultaneously acquired untreated *E. coli* 2452.

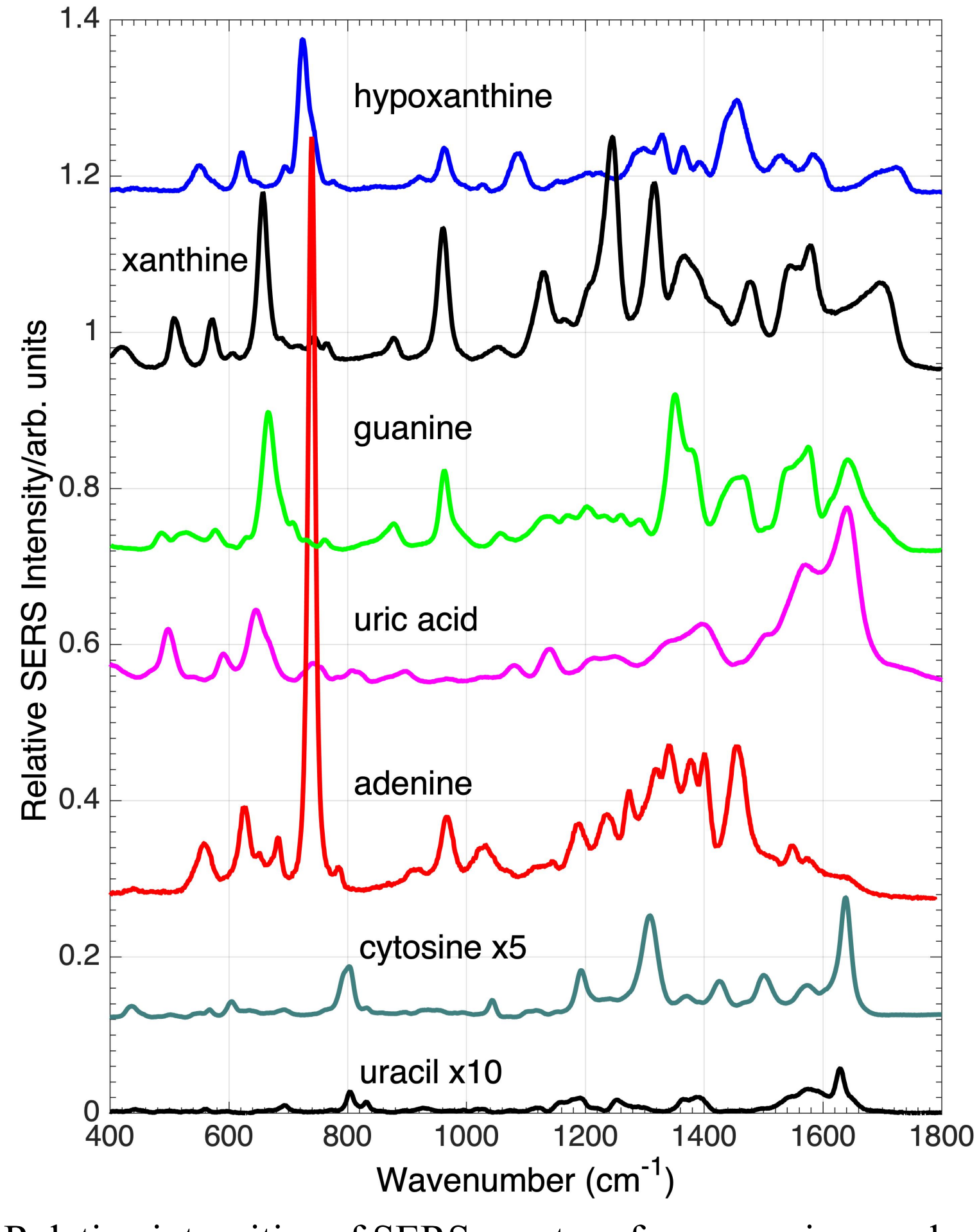

Fig. S16. Relative intensities of SERS spectra of some purines and pyrimidine on Au SERS substrates excited at 785 nm.